\documentclass[12pt]{article}

\usepackage{color}
\usepackage{graphicx}
\usepackage{hyperref}

\newcommand{\cD}{{\cal{D}}}

\newcommand{\cL}{{\cal{L}}}
\newcommand{\cN}{{\cal{N}}}

\newcommand{\bld}[1]{\mbox{\boldmath $#1$}}

\newcommand{\bbc}{\bld c}
\newcommand{\bbd}{\bld d}
\newcommand{\bbe}{\bld e}

\newcommand{\bbg}{\bld g}

\newcommand{\bbs}{\bld s}

\newcommand{\bbu}{\bld u}
\newcommand{\bbv}{\bld v}
\newcommand{\bbw}{\bld w}
\newcommand{\bbx}{\bld x}
\newcommand{\bby}{\bld y}

\newcommand{\bbC}{\bld C}

\newcommand{\bbQ}{\bld Q}

\newcommand{\bbY}{\bld Y}

\definecolor{white}{rgb}{0.996,0.996,0.996}
\definecolor{aliceblue}{rgb}{0.938,0.969,0.996}
\definecolor{antiquewhite}{rgb}{0.977,0.918,0.84}
\definecolor{antiquewhite1}{rgb}{0.996,0.934,0.855}
\definecolor{antiquewhite2}{rgb}{0.93,0.871,0.797}
\definecolor{antiquewhite3}{rgb}{0.801,0.75,0.688}
\definecolor{antiquewhite4}{rgb}{0.543,0.512,0.469}
\definecolor{aquamarine}{rgb}{0.496,0.996,0.828}
\definecolor{aquamarine1}{rgb}{0.496,0.996,0.828}
\definecolor{aquamarine2}{rgb}{0.461,0.93,0.773}
\definecolor{aquamarine3}{rgb}{0.398,0.801,0.664}
\definecolor{aquamarine4}{rgb}{0.27,0.543,0.453}
\definecolor{azure}{rgb}{0.938,0.996,0.996}
\definecolor{azure1}{rgb}{0.938,0.996,0.996}
\definecolor{azure2}{rgb}{0.875,0.93,0.93}
\definecolor{azure3}{rgb}{0.754,0.801,0.801}
\definecolor{azure4}{rgb}{0.512,0.543,0.543}
\definecolor{beige}{rgb}{0.957,0.957,0.859}
\definecolor{bisque}{rgb}{0.996,0.891,0.766}
\definecolor{bisque1}{rgb}{0.996,0.891,0.766}
\definecolor{bisque2}{rgb}{0.93,0.832,0.715}
\definecolor{bisque3}{rgb}{0.801,0.715,0.617}
\definecolor{bisque4}{rgb}{0.543,0.488,0.418}
\definecolor{black}{rgb}{0,0,0}
\definecolor{blanchedalmond}{rgb}{0.996,0.918,0.801}
\definecolor{blue}{rgb}{0,0,0.996}
\definecolor{blue1}{rgb}{0,0,0.996}
\definecolor{blue2}{rgb}{0,0,0.93}
\definecolor{blue3}{rgb}{0,0,0.801}
\definecolor{blue4}{rgb}{0,0,0.543}
\definecolor{blueviolet}{rgb}{0.539,0.168,0.883}
\definecolor{brown}{rgb}{0.645,0.164,0.164}
\definecolor{brown1}{rgb}{0.996,0.25,0.25}
\definecolor{brown2}{rgb}{0.93,0.23,0.23}
\definecolor{brown3}{rgb}{0.801,0.199,0.199}
\definecolor{brown4}{rgb}{0.543,0.137,0.137}
\definecolor{burlywood}{rgb}{0.867,0.719,0.527}
\definecolor{burlywood1}{rgb}{0.996,0.824,0.605}
\definecolor{burlywood2}{rgb}{0.93,0.77,0.566}
\definecolor{burlywood3}{rgb}{0.801,0.664,0.488}
\definecolor{burlywood4}{rgb}{0.543,0.449,0.332}
\definecolor{cadetblue}{rgb}{0.371,0.617,0.625}
\definecolor{cadetblue1}{rgb}{0.594,0.957,0.996}
\definecolor{cadetblue2}{rgb}{0.555,0.895,0.93}
\definecolor{cadetblue3}{rgb}{0.477,0.77,0.801}
\definecolor{cadetblue4}{rgb}{0.324,0.523,0.543}
\definecolor{chartreuse}{rgb}{0.496,0.996,0}
\definecolor{chartreuse1}{rgb}{0.496,0.996,0}
\definecolor{chartreuse2}{rgb}{0.461,0.93,0}
\definecolor{chartreuse3}{rgb}{0.398,0.801,0}
\definecolor{chartreuse4}{rgb}{0.27,0.543,0}
\definecolor{chocolate}{rgb}{0.82,0.41,0.117}
\definecolor{chocolate1}{rgb}{0.996,0.496,0.141}
\definecolor{chocolate2}{rgb}{0.93,0.461,0.129}
\definecolor{chocolate3}{rgb}{0.801,0.398,0.113}
\definecolor{chocolate4}{rgb}{0.543,0.27,0.0742}
\definecolor{coral}{rgb}{0.996,0.496,0.312}
\definecolor{coral1}{rgb}{0.996,0.445,0.336}
\definecolor{coral2}{rgb}{0.93,0.414,0.312}
\definecolor{coral3}{rgb}{0.801,0.355,0.27}
\definecolor{coral4}{rgb}{0.543,0.242,0.184}
\definecolor{cornflowerblue}{rgb}{0.391,0.582,0.926}
\definecolor{cornsilk}{rgb}{0.996,0.969,0.859}
\definecolor{cornsilk1}{rgb}{0.996,0.969,0.859}
\definecolor{cornsilk2}{rgb}{0.93,0.906,0.801}
\definecolor{cornsilk3}{rgb}{0.801,0.781,0.691}
\definecolor{cornsilk4}{rgb}{0.543,0.531,0.469}
\definecolor{cyan}{rgb}{0,0.996,0.996}
\definecolor{cyan1}{rgb}{0,0.996,0.996}
\definecolor{cyan2}{rgb}{0,0.93,0.93}
\definecolor{cyan3}{rgb}{0,0.801,0.801}
\definecolor{cyan4}{rgb}{0,0.543,0.543}
\definecolor{darkblue}{rgb}{0,0,0.543}
\definecolor{darkcyan}{rgb}{0,0.543,0.543}
\definecolor{darkgoldenrod}{rgb}{0.719,0.523,0.043}
\definecolor{darkgoldenrod1}{rgb}{0.996,0.723,0.0586}
\definecolor{darkgoldenrod2}{rgb}{0.93,0.676,0.0547}
\definecolor{darkgoldenrod3}{rgb}{0.801,0.582,0.0469}
\definecolor{darkgoldenrod4}{rgb}{0.543,0.395,0.0312}
\definecolor{darkgray}{rgb}{0.66,0.66,0.66}
\definecolor{darkgreen}{rgb}{0,0.391,0}
\definecolor{darkgrey}{rgb}{0.66,0.66,0.66}
\definecolor{darkkhaki}{rgb}{0.738,0.715,0.418}
\definecolor{darkmagenta}{rgb}{0.543,0,0.543}
\definecolor{darkolivegreen}{rgb}{0.332,0.418,0.184}
\definecolor{darkolivegreen1}{rgb}{0.789,0.996,0.438}
\definecolor{darkolivegreen2}{rgb}{0.734,0.93,0.406}
\definecolor{darkolivegreen3}{rgb}{0.633,0.801,0.352}
\definecolor{darkolivegreen4}{rgb}{0.43,0.543,0.238}
\definecolor{darkorange}{rgb}{0.996,0.547,0}
\definecolor{darkorange1}{rgb}{0.996,0.496,0}
\definecolor{darkorange2}{rgb}{0.93,0.461,0}
\definecolor{darkorange3}{rgb}{0.801,0.398,0}
\definecolor{darkorange4}{rgb}{0.543,0.27,0}
\definecolor{darkorchid}{rgb}{0.598,0.195,0.797}
\definecolor{darkorchid1}{rgb}{0.746,0.242,0.996}
\definecolor{darkorchid2}{rgb}{0.695,0.227,0.93}
\definecolor{darkorchid3}{rgb}{0.602,0.195,0.801}
\definecolor{darkorchid4}{rgb}{0.406,0.133,0.543}
\definecolor{darkred}{rgb}{0.543,0,0}
\definecolor{darksalmon}{rgb}{0.91,0.586,0.477}
\definecolor{darkseagreen}{rgb}{0.559,0.734,0.559}
\definecolor{darkseagreen1}{rgb}{0.754,0.996,0.754}
\definecolor{darkseagreen2}{rgb}{0.703,0.93,0.703}
\definecolor{darkseagreen3}{rgb}{0.605,0.801,0.605}
\definecolor{darkseagreen4}{rgb}{0.41,0.543,0.41}
\definecolor{darkslateblue}{rgb}{0.281,0.238,0.543}
\definecolor{darkslategray}{rgb}{0.184,0.309,0.309}
\definecolor{darkslategray1}{rgb}{0.59,0.996,0.996}
\definecolor{darkslategray2}{rgb}{0.551,0.93,0.93}
\definecolor{darkslategray3}{rgb}{0.473,0.801,0.801}
\definecolor{darkslategray4}{rgb}{0.32,0.543,0.543}
\definecolor{darkslategrey}{rgb}{0.184,0.309,0.309}
\definecolor{darkturquoise}{rgb}{0,0.805,0.816}
\definecolor{darkviolet}{rgb}{0.578,0,0.824}
\definecolor{deeppink}{rgb}{0.996,0.0781,0.574}
\definecolor{deeppink1}{rgb}{0.996,0.0781,0.574}
\definecolor{deeppink2}{rgb}{0.93,0.0703,0.535}
\definecolor{deeppink3}{rgb}{0.801,0.0625,0.461}
\definecolor{deeppink4}{rgb}{0.543,0.0391,0.312}
\definecolor{deepskyblue}{rgb}{0,0.746,0.996}
\definecolor{deepskyblue1}{rgb}{0,0.746,0.996}
\definecolor{deepskyblue2}{rgb}{0,0.695,0.93}
\definecolor{deepskyblue3}{rgb}{0,0.602,0.801}
\definecolor{deepskyblue4}{rgb}{0,0.406,0.543}
\definecolor{dimgray}{rgb}{0.41,0.41,0.41}
\definecolor{dimgrey}{rgb}{0.41,0.41,0.41}
\definecolor{dodgerblue}{rgb}{0.117,0.562,0.996}
\definecolor{dodgerblue1}{rgb}{0.117,0.562,0.996}
\definecolor{dodgerblue2}{rgb}{0.109,0.523,0.93}
\definecolor{dodgerblue3}{rgb}{0.0938,0.453,0.801}
\definecolor{dodgerblue4}{rgb}{0.0625,0.305,0.543}
\definecolor{firebrick}{rgb}{0.695,0.133,0.133}
\definecolor{firebrick1}{rgb}{0.996,0.188,0.188}
\definecolor{firebrick2}{rgb}{0.93,0.172,0.172}
\definecolor{firebrick3}{rgb}{0.801,0.148,0.148}
\definecolor{firebrick4}{rgb}{0.543,0.102,0.102}
\definecolor{floralwhite}{rgb}{0.996,0.977,0.938}
\definecolor{forestgreen}{rgb}{0.133,0.543,0.133}
\definecolor{gainsboro}{rgb}{0.859,0.859,0.859}
\definecolor{ghostwhite}{rgb}{0.969,0.969,0.996}
\definecolor{gold}{rgb}{0.996,0.84,0}
\definecolor{gold1}{rgb}{0.996,0.84,0}
\definecolor{gold2}{rgb}{0.93,0.785,0}
\definecolor{gold3}{rgb}{0.801,0.676,0}
\definecolor{gold4}{rgb}{0.543,0.457,0}
\definecolor{goldenrod}{rgb}{0.852,0.645,0.125}
\definecolor{goldenrod1}{rgb}{0.996,0.754,0.145}
\definecolor{goldenrod2}{rgb}{0.93,0.703,0.133}
\definecolor{goldenrod3}{rgb}{0.801,0.605,0.113}
\definecolor{goldenrod4}{rgb}{0.543,0.41,0.0781}
\definecolor{gray}{rgb}{0.742,0.742,0.742}
\definecolor{gray0}{rgb}{0,0,0}
\definecolor{gray1}{rgb}{0.0117,0.0117,0.0117}
\definecolor{gray2}{rgb}{0.0195,0.0195,0.0195}
\definecolor{gray3}{rgb}{0.0312,0.0312,0.0312}
\definecolor{gray4}{rgb}{0.0391,0.0391,0.0391}
\definecolor{gray5}{rgb}{0.0508,0.0508,0.0508}
\definecolor{gray6}{rgb}{0.0586,0.0586,0.0586}
\definecolor{gray7}{rgb}{0.0703,0.0703,0.0703}
\definecolor{gray8}{rgb}{0.0781,0.0781,0.0781}
\definecolor{gray9}{rgb}{0.0898,0.0898,0.0898}
\definecolor{gray10}{rgb}{0.102,0.102,0.102}
\definecolor{gray11}{rgb}{0.109,0.109,0.109}
\definecolor{gray12}{rgb}{0.121,0.121,0.121}
\definecolor{gray13}{rgb}{0.129,0.129,0.129}
\definecolor{gray14}{rgb}{0.141,0.141,0.141}
\definecolor{gray15}{rgb}{0.148,0.148,0.148}
\definecolor{gray16}{rgb}{0.16,0.16,0.16}
\definecolor{gray17}{rgb}{0.168,0.168,0.168}
\definecolor{gray18}{rgb}{0.18,0.18,0.18}
\definecolor{gray19}{rgb}{0.188,0.188,0.188}
\definecolor{gray20}{rgb}{0.199,0.199,0.199}
\definecolor{gray21}{rgb}{0.211,0.211,0.211}
\definecolor{gray22}{rgb}{0.219,0.219,0.219}
\definecolor{gray23}{rgb}{0.23,0.23,0.23}
\definecolor{gray24}{rgb}{0.238,0.238,0.238}
\definecolor{gray25}{rgb}{0.25,0.25,0.25}
\definecolor{gray26}{rgb}{0.258,0.258,0.258}
\definecolor{gray27}{rgb}{0.27,0.27,0.27}
\definecolor{gray28}{rgb}{0.277,0.277,0.277}
\definecolor{gray29}{rgb}{0.289,0.289,0.289}
\definecolor{gray30}{rgb}{0.301,0.301,0.301}
\definecolor{gray31}{rgb}{0.309,0.309,0.309}
\definecolor{gray32}{rgb}{0.32,0.32,0.32}
\definecolor{gray33}{rgb}{0.328,0.328,0.328}
\definecolor{gray34}{rgb}{0.34,0.34,0.34}
\definecolor{gray35}{rgb}{0.348,0.348,0.348}
\definecolor{gray36}{rgb}{0.359,0.359,0.359}
\definecolor{gray37}{rgb}{0.367,0.367,0.367}
\definecolor{gray38}{rgb}{0.379,0.379,0.379}
\definecolor{gray39}{rgb}{0.387,0.387,0.387}
\definecolor{gray40}{rgb}{0.398,0.398,0.398}
\definecolor{gray41}{rgb}{0.41,0.41,0.41}
\definecolor{gray42}{rgb}{0.418,0.418,0.418}
\definecolor{gray43}{rgb}{0.43,0.43,0.43}
\definecolor{gray44}{rgb}{0.438,0.438,0.438}
\definecolor{gray45}{rgb}{0.449,0.449,0.449}
\definecolor{gray46}{rgb}{0.457,0.457,0.457}
\definecolor{gray47}{rgb}{0.469,0.469,0.469}
\definecolor{gray48}{rgb}{0.477,0.477,0.477}
\definecolor{gray49}{rgb}{0.488,0.488,0.488}
\definecolor{gray50}{rgb}{0.496,0.496,0.496}
\definecolor{gray51}{rgb}{0.508,0.508,0.508}
\definecolor{gray52}{rgb}{0.52,0.52,0.52}
\definecolor{gray53}{rgb}{0.527,0.527,0.527}
\definecolor{gray54}{rgb}{0.539,0.539,0.539}
\definecolor{gray55}{rgb}{0.547,0.547,0.547}
\definecolor{gray56}{rgb}{0.559,0.559,0.559}
\definecolor{gray57}{rgb}{0.566,0.566,0.566}
\definecolor{gray58}{rgb}{0.578,0.578,0.578}
\definecolor{gray59}{rgb}{0.586,0.586,0.586}
\definecolor{gray60}{rgb}{0.598,0.598,0.598}
\definecolor{gray61}{rgb}{0.609,0.609,0.609}
\definecolor{gray62}{rgb}{0.617,0.617,0.617}
\definecolor{gray63}{rgb}{0.629,0.629,0.629}
\definecolor{gray64}{rgb}{0.637,0.637,0.637}
\definecolor{gray65}{rgb}{0.648,0.648,0.648}
\definecolor{gray66}{rgb}{0.656,0.656,0.656}
\definecolor{gray67}{rgb}{0.668,0.668,0.668}
\definecolor{gray68}{rgb}{0.676,0.676,0.676}
\definecolor{gray69}{rgb}{0.688,0.688,0.688}
\definecolor{gray70}{rgb}{0.699,0.699,0.699}
\definecolor{gray71}{rgb}{0.707,0.707,0.707}
\definecolor{gray72}{rgb}{0.719,0.719,0.719}
\definecolor{gray73}{rgb}{0.727,0.727,0.727}
\definecolor{gray74}{rgb}{0.738,0.738,0.738}
\definecolor{gray75}{rgb}{0.746,0.746,0.746}
\definecolor{gray76}{rgb}{0.758,0.758,0.758}
\definecolor{gray77}{rgb}{0.766,0.766,0.766}
\definecolor{gray78}{rgb}{0.777,0.777,0.777}
\definecolor{gray79}{rgb}{0.785,0.785,0.785}
\definecolor{gray80}{rgb}{0.797,0.797,0.797}
\definecolor{gray81}{rgb}{0.809,0.809,0.809}
\definecolor{gray82}{rgb}{0.816,0.816,0.816}
\definecolor{gray83}{rgb}{0.828,0.828,0.828}
\definecolor{gray84}{rgb}{0.836,0.836,0.836}
\definecolor{gray85}{rgb}{0.848,0.848,0.848}
\definecolor{gray86}{rgb}{0.855,0.855,0.855}
\definecolor{gray87}{rgb}{0.867,0.867,0.867}
\definecolor{gray88}{rgb}{0.875,0.875,0.875}
\definecolor{gray89}{rgb}{0.887,0.887,0.887}
\definecolor{gray90}{rgb}{0.895,0.895,0.895}
\definecolor{gray91}{rgb}{0.906,0.906,0.906}
\definecolor{gray92}{rgb}{0.918,0.918,0.918}
\definecolor{gray93}{rgb}{0.926,0.926,0.926}
\definecolor{gray94}{rgb}{0.938,0.938,0.938}
\definecolor{gray95}{rgb}{0.945,0.945,0.945}
\definecolor{gray96}{rgb}{0.957,0.957,0.957}
\definecolor{gray97}{rgb}{0.965,0.965,0.965}
\definecolor{gray98}{rgb}{0.977,0.977,0.977}
\definecolor{gray99}{rgb}{0.984,0.984,0.984}
\definecolor{gray100}{rgb}{0.996,0.996,0.996}
\definecolor{green}{rgb}{0,0.996,0}
\definecolor{green1}{rgb}{0,0.996,0}
\definecolor{green2}{rgb}{0,0.93,0}
\definecolor{green3}{rgb}{0,0.801,0}
\definecolor{green4}{rgb}{0,0.543,0}
\definecolor{greenyellow}{rgb}{0.676,0.996,0.184}
\definecolor{grey}{rgb}{0.742,0.742,0.742}
\definecolor{grey0}{rgb}{0,0,0}
\definecolor{grey1}{rgb}{0.0117,0.0117,0.0117}
\definecolor{grey2}{rgb}{0.0195,0.0195,0.0195}
\definecolor{grey3}{rgb}{0.0312,0.0312,0.0312}
\definecolor{grey4}{rgb}{0.0391,0.0391,0.0391}
\definecolor{grey5}{rgb}{0.0508,0.0508,0.0508}
\definecolor{grey6}{rgb}{0.0586,0.0586,0.0586}
\definecolor{grey7}{rgb}{0.0703,0.0703,0.0703}
\definecolor{grey8}{rgb}{0.0781,0.0781,0.0781}
\definecolor{grey9}{rgb}{0.0898,0.0898,0.0898}
\definecolor{grey10}{rgb}{0.102,0.102,0.102}
\definecolor{grey11}{rgb}{0.109,0.109,0.109}
\definecolor{grey12}{rgb}{0.121,0.121,0.121}
\definecolor{grey13}{rgb}{0.129,0.129,0.129}
\definecolor{grey14}{rgb}{0.141,0.141,0.141}
\definecolor{grey15}{rgb}{0.148,0.148,0.148}
\definecolor{grey16}{rgb}{0.16,0.16,0.16}
\definecolor{grey17}{rgb}{0.168,0.168,0.168}
\definecolor{grey18}{rgb}{0.18,0.18,0.18}
\definecolor{grey19}{rgb}{0.188,0.188,0.188}
\definecolor{grey20}{rgb}{0.199,0.199,0.199}
\definecolor{grey21}{rgb}{0.211,0.211,0.211}
\definecolor{grey22}{rgb}{0.219,0.219,0.219}
\definecolor{grey23}{rgb}{0.23,0.23,0.23}
\definecolor{grey24}{rgb}{0.238,0.238,0.238}
\definecolor{grey25}{rgb}{0.25,0.25,0.25}
\definecolor{grey26}{rgb}{0.258,0.258,0.258}
\definecolor{grey27}{rgb}{0.27,0.27,0.27}
\definecolor{grey28}{rgb}{0.277,0.277,0.277}
\definecolor{grey29}{rgb}{0.289,0.289,0.289}
\definecolor{grey30}{rgb}{0.301,0.301,0.301}
\definecolor{grey31}{rgb}{0.309,0.309,0.309}
\definecolor{grey32}{rgb}{0.32,0.32,0.32}
\definecolor{grey33}{rgb}{0.328,0.328,0.328}
\definecolor{grey34}{rgb}{0.34,0.34,0.34}
\definecolor{grey35}{rgb}{0.348,0.348,0.348}
\definecolor{grey36}{rgb}{0.359,0.359,0.359}
\definecolor{grey37}{rgb}{0.367,0.367,0.367}
\definecolor{grey38}{rgb}{0.379,0.379,0.379}
\definecolor{grey39}{rgb}{0.387,0.387,0.387}
\definecolor{grey40}{rgb}{0.398,0.398,0.398}
\definecolor{grey41}{rgb}{0.41,0.41,0.41}
\definecolor{grey42}{rgb}{0.418,0.418,0.418}
\definecolor{grey43}{rgb}{0.43,0.43,0.43}
\definecolor{grey44}{rgb}{0.438,0.438,0.438}
\definecolor{grey45}{rgb}{0.449,0.449,0.449}
\definecolor{grey46}{rgb}{0.457,0.457,0.457}
\definecolor{grey47}{rgb}{0.469,0.469,0.469}
\definecolor{grey48}{rgb}{0.477,0.477,0.477}
\definecolor{grey49}{rgb}{0.488,0.488,0.488}
\definecolor{grey50}{rgb}{0.496,0.496,0.496}
\definecolor{grey51}{rgb}{0.508,0.508,0.508}
\definecolor{grey52}{rgb}{0.52,0.52,0.52}
\definecolor{grey53}{rgb}{0.527,0.527,0.527}
\definecolor{grey54}{rgb}{0.539,0.539,0.539}
\definecolor{grey55}{rgb}{0.547,0.547,0.547}
\definecolor{grey56}{rgb}{0.559,0.559,0.559}
\definecolor{grey57}{rgb}{0.566,0.566,0.566}
\definecolor{grey58}{rgb}{0.578,0.578,0.578}
\definecolor{grey59}{rgb}{0.586,0.586,0.586}
\definecolor{grey60}{rgb}{0.598,0.598,0.598}
\definecolor{grey61}{rgb}{0.609,0.609,0.609}
\definecolor{grey62}{rgb}{0.617,0.617,0.617}
\definecolor{grey63}{rgb}{0.629,0.629,0.629}
\definecolor{grey64}{rgb}{0.637,0.637,0.637}
\definecolor{grey65}{rgb}{0.648,0.648,0.648}
\definecolor{grey66}{rgb}{0.656,0.656,0.656}
\definecolor{grey67}{rgb}{0.668,0.668,0.668}
\definecolor{grey68}{rgb}{0.676,0.676,0.676}
\definecolor{grey69}{rgb}{0.688,0.688,0.688}
\definecolor{grey70}{rgb}{0.699,0.699,0.699}
\definecolor{grey71}{rgb}{0.707,0.707,0.707}
\definecolor{grey72}{rgb}{0.719,0.719,0.719}
\definecolor{grey73}{rgb}{0.727,0.727,0.727}
\definecolor{grey74}{rgb}{0.738,0.738,0.738}
\definecolor{grey75}{rgb}{0.746,0.746,0.746}
\definecolor{grey76}{rgb}{0.758,0.758,0.758}
\definecolor{grey77}{rgb}{0.766,0.766,0.766}
\definecolor{grey78}{rgb}{0.777,0.777,0.777}
\definecolor{grey79}{rgb}{0.785,0.785,0.785}
\definecolor{grey80}{rgb}{0.797,0.797,0.797}
\definecolor{grey81}{rgb}{0.809,0.809,0.809}
\definecolor{grey82}{rgb}{0.816,0.816,0.816}
\definecolor{grey83}{rgb}{0.828,0.828,0.828}
\definecolor{grey84}{rgb}{0.836,0.836,0.836}
\definecolor{grey85}{rgb}{0.848,0.848,0.848}
\definecolor{grey86}{rgb}{0.855,0.855,0.855}
\definecolor{grey87}{rgb}{0.867,0.867,0.867}
\definecolor{grey88}{rgb}{0.875,0.875,0.875}
\definecolor{grey89}{rgb}{0.887,0.887,0.887}
\definecolor{grey90}{rgb}{0.895,0.895,0.895}
\definecolor{grey91}{rgb}{0.906,0.906,0.906}
\definecolor{grey92}{rgb}{0.918,0.918,0.918}
\definecolor{grey93}{rgb}{0.926,0.926,0.926}
\definecolor{grey94}{rgb}{0.938,0.938,0.938}
\definecolor{grey95}{rgb}{0.945,0.945,0.945}
\definecolor{grey96}{rgb}{0.957,0.957,0.957}
\definecolor{grey97}{rgb}{0.965,0.965,0.965}
\definecolor{grey98}{rgb}{0.977,0.977,0.977}
\definecolor{grey99}{rgb}{0.984,0.984,0.984}
\definecolor{grey100}{rgb}{0.996,0.996,0.996}
\definecolor{honeydew}{rgb}{0.938,0.996,0.938}
\definecolor{honeydew1}{rgb}{0.938,0.996,0.938}
\definecolor{honeydew2}{rgb}{0.875,0.93,0.875}
\definecolor{honeydew3}{rgb}{0.754,0.801,0.754}
\definecolor{honeydew4}{rgb}{0.512,0.543,0.512}
\definecolor{hotpink}{rgb}{0.996,0.41,0.703}
\definecolor{hotpink1}{rgb}{0.996,0.43,0.703}
\definecolor{hotpink2}{rgb}{0.93,0.414,0.652}
\definecolor{hotpink3}{rgb}{0.801,0.375,0.562}
\definecolor{hotpink4}{rgb}{0.543,0.227,0.383}
\definecolor{indianred}{rgb}{0.801,0.359,0.359}
\definecolor{indianred1}{rgb}{0.996,0.414,0.414}
\definecolor{indianred2}{rgb}{0.93,0.387,0.387}
\definecolor{indianred3}{rgb}{0.801,0.332,0.332}
\definecolor{indianred4}{rgb}{0.543,0.227,0.227}
\definecolor{ivory}{rgb}{0.996,0.996,0.938}
\definecolor{ivory1}{rgb}{0.996,0.996,0.938}
\definecolor{ivory2}{rgb}{0.93,0.93,0.875}
\definecolor{ivory3}{rgb}{0.801,0.801,0.754}
\definecolor{ivory4}{rgb}{0.543,0.543,0.512}
\definecolor{khaki}{rgb}{0.938,0.898,0.547}
\definecolor{khaki1}{rgb}{0.996,0.961,0.559}
\definecolor{khaki2}{rgb}{0.93,0.898,0.52}
\definecolor{khaki3}{rgb}{0.801,0.773,0.449}
\definecolor{khaki4}{rgb}{0.543,0.523,0.305}
\definecolor{lavender}{rgb}{0.898,0.898,0.977}
\definecolor{lavenderblush}{rgb}{0.996,0.938,0.957}
\definecolor{lavenderblush1}{rgb}{0.996,0.938,0.957}
\definecolor{lavenderblush2}{rgb}{0.93,0.875,0.895}
\definecolor{lavenderblush3}{rgb}{0.801,0.754,0.77}
\definecolor{lavenderblush4}{rgb}{0.543,0.512,0.523}
\definecolor{lawngreen}{rgb}{0.484,0.984,0}
\definecolor{lemonchiffon}{rgb}{0.996,0.977,0.801}
\definecolor{lemonchiffon1}{rgb}{0.996,0.977,0.801}
\definecolor{lemonchiffon2}{rgb}{0.93,0.91,0.746}
\definecolor{lemonchiffon3}{rgb}{0.801,0.785,0.645}
\definecolor{lemonchiffon4}{rgb}{0.543,0.535,0.438}
\definecolor{lightblue}{rgb}{0.676,0.844,0.898}
\definecolor{lightblue1}{rgb}{0.746,0.934,0.996}
\definecolor{lightblue2}{rgb}{0.695,0.871,0.93}
\definecolor{lightblue3}{rgb}{0.602,0.75,0.801}
\definecolor{lightblue4}{rgb}{0.406,0.512,0.543}
\definecolor{lightcoral}{rgb}{0.938,0.5,0.5}
\definecolor{lightcyan}{rgb}{0.875,0.996,0.996}
\definecolor{lightcyan1}{rgb}{0.875,0.996,0.996}
\definecolor{lightcyan2}{rgb}{0.816,0.93,0.93}
\definecolor{lightcyan3}{rgb}{0.703,0.801,0.801}
\definecolor{lightcyan4}{rgb}{0.477,0.543,0.543}
\definecolor{lightgoldenrod}{rgb}{0.93,0.863,0.508}
\definecolor{lightgoldenrod1}{rgb}{0.996,0.922,0.543}
\definecolor{lightgoldenrod2}{rgb}{0.93,0.859,0.508}
\definecolor{lightgoldenrod3}{rgb}{0.801,0.742,0.438}
\definecolor{lightgoldenrod4}{rgb}{0.543,0.504,0.297}
\definecolor{lightgoldenrodyellow}{rgb}{0.977,0.977,0.82}
\definecolor{lightgray}{rgb}{0.824,0.824,0.824}
\definecolor{lightgreen}{rgb}{0.562,0.93,0.562}
\definecolor{lightgrey}{rgb}{0.824,0.824,0.824}
\definecolor{lightpink}{rgb}{0.996,0.711,0.754}
\definecolor{lightpink1}{rgb}{0.996,0.68,0.723}
\definecolor{lightpink2}{rgb}{0.93,0.633,0.676}
\definecolor{lightpink3}{rgb}{0.801,0.547,0.582}
\definecolor{lightpink4}{rgb}{0.543,0.371,0.395}
\definecolor{lightsalmon}{rgb}{0.996,0.625,0.477}
\definecolor{lightsalmon1}{rgb}{0.996,0.625,0.477}
\definecolor{lightsalmon2}{rgb}{0.93,0.582,0.445}
\definecolor{lightsalmon3}{rgb}{0.801,0.504,0.383}
\definecolor{lightsalmon4}{rgb}{0.543,0.34,0.258}
\definecolor{lightseagreen}{rgb}{0.125,0.695,0.664}
\definecolor{lightskyblue}{rgb}{0.527,0.805,0.977}
\definecolor{lightskyblue1}{rgb}{0.688,0.883,0.996}
\definecolor{lightskyblue2}{rgb}{0.641,0.824,0.93}
\definecolor{lightskyblue3}{rgb}{0.551,0.711,0.801}
\definecolor{lightskyblue4}{rgb}{0.375,0.48,0.543}
\definecolor{lightslateblue}{rgb}{0.516,0.438,0.996}
\definecolor{lightslategray}{rgb}{0.465,0.531,0.598}
\definecolor{lightslategrey}{rgb}{0.465,0.531,0.598}
\definecolor{lightsteelblue}{rgb}{0.688,0.766,0.867}
\definecolor{lightsteelblue1}{rgb}{0.789,0.879,0.996}
\definecolor{lightsteelblue2}{rgb}{0.734,0.82,0.93}
\definecolor{lightsteelblue3}{rgb}{0.633,0.707,0.801}
\definecolor{lightsteelblue4}{rgb}{0.43,0.48,0.543}
\definecolor{lightyellow}{rgb}{0.996,0.996,0.875}
\definecolor{lightyellow1}{rgb}{0.996,0.996,0.875}
\definecolor{lightyellow2}{rgb}{0.93,0.93,0.816}
\definecolor{lightyellow3}{rgb}{0.801,0.801,0.703}
\definecolor{lightyellow4}{rgb}{0.543,0.543,0.477}
\definecolor{limegreen}{rgb}{0.195,0.801,0.195}
\definecolor{linen}{rgb}{0.977,0.938,0.898}
\definecolor{magenta}{rgb}{0.996,0,0.996}
\definecolor{magenta1}{rgb}{0.996,0,0.996}
\definecolor{magenta2}{rgb}{0.93,0,0.93}
\definecolor{magenta3}{rgb}{0.801,0,0.801}
\definecolor{magenta4}{rgb}{0.543,0,0.543}
\definecolor{maroon}{rgb}{0.688,0.188,0.375}
\definecolor{maroon1}{rgb}{0.996,0.203,0.699}
\definecolor{maroon2}{rgb}{0.93,0.188,0.652}
\definecolor{maroon3}{rgb}{0.801,0.16,0.562}
\definecolor{maroon4}{rgb}{0.543,0.109,0.383}
\definecolor{mediumaquamarine}{rgb}{0.398,0.801,0.664}
\definecolor{mediumblue}{rgb}{0,0,0.801}
\definecolor{mediumorchid}{rgb}{0.727,0.332,0.824}
\definecolor{mediumorchid1}{rgb}{0.875,0.398,0.996}
\definecolor{mediumorchid2}{rgb}{0.816,0.371,0.93}
\definecolor{mediumorchid3}{rgb}{0.703,0.32,0.801}
\definecolor{mediumorchid4}{rgb}{0.477,0.215,0.543}
\definecolor{mediumpurple}{rgb}{0.574,0.438,0.855}
\definecolor{mediumpurple1}{rgb}{0.668,0.508,0.996}
\definecolor{mediumpurple2}{rgb}{0.621,0.473,0.93}
\definecolor{mediumpurple3}{rgb}{0.535,0.406,0.801}
\definecolor{mediumpurple4}{rgb}{0.363,0.277,0.543}
\definecolor{mediumseagreen}{rgb}{0.234,0.699,0.441}
\definecolor{mediumslateblue}{rgb}{0.48,0.406,0.93}
\definecolor{mediumspringgreen}{rgb}{0,0.977,0.602}
\definecolor{mediumturquoise}{rgb}{0.281,0.816,0.797}
\definecolor{mediumvioletred}{rgb}{0.777,0.082,0.52}
\definecolor{midnightblue}{rgb}{0.0977,0.0977,0.438}
\definecolor{mintcream}{rgb}{0.957,0.996,0.977}
\definecolor{mistyrose}{rgb}{0.996,0.891,0.879}
\definecolor{mistyrose1}{rgb}{0.996,0.891,0.879}
\definecolor{mistyrose2}{rgb}{0.93,0.832,0.82}
\definecolor{mistyrose3}{rgb}{0.801,0.715,0.707}
\definecolor{mistyrose4}{rgb}{0.543,0.488,0.48}
\definecolor{moccasin}{rgb}{0.996,0.891,0.707}
\definecolor{navajowhite}{rgb}{0.996,0.867,0.676}
\definecolor{navajowhite1}{rgb}{0.996,0.867,0.676}
\definecolor{navajowhite2}{rgb}{0.93,0.809,0.629}
\definecolor{navajowhite3}{rgb}{0.801,0.699,0.543}
\definecolor{navajowhite4}{rgb}{0.543,0.473,0.367}
\definecolor{navy}{rgb}{0,0,0.5}
\definecolor{navyblue}{rgb}{0,0,0.5}
\definecolor{oldlace}{rgb}{0.988,0.957,0.898}
\definecolor{olivedrab}{rgb}{0.418,0.555,0.137}
\definecolor{olivedrab1}{rgb}{0.75,0.996,0.242}
\definecolor{olivedrab2}{rgb}{0.699,0.93,0.227}
\definecolor{olivedrab3}{rgb}{0.602,0.801,0.195}
\definecolor{olivedrab4}{rgb}{0.41,0.543,0.133}
\definecolor{orange}{rgb}{0.996,0.645,0}
\definecolor{orange1}{rgb}{0.996,0.645,0}
\definecolor{orange2}{rgb}{0.93,0.602,0}
\definecolor{orange3}{rgb}{0.801,0.52,0}
\definecolor{orange4}{rgb}{0.543,0.352,0}
\definecolor{orangered}{rgb}{0.996,0.27,0}
\definecolor{orangered1}{rgb}{0.996,0.27,0}
\definecolor{orangered2}{rgb}{0.93,0.25,0}
\definecolor{orangered3}{rgb}{0.801,0.215,0}
\definecolor{orangered4}{rgb}{0.543,0.145,0}
\definecolor{orchid}{rgb}{0.852,0.438,0.836}
\definecolor{orchid1}{rgb}{0.996,0.512,0.977}
\definecolor{orchid2}{rgb}{0.93,0.477,0.91}
\definecolor{orchid3}{rgb}{0.801,0.41,0.785}
\definecolor{orchid4}{rgb}{0.543,0.277,0.535}
\definecolor{palegoldenrod}{rgb}{0.93,0.906,0.664}
\definecolor{palegreen}{rgb}{0.594,0.98,0.594}
\definecolor{palegreen1}{rgb}{0.602,0.996,0.602}
\definecolor{palegreen2}{rgb}{0.562,0.93,0.562}
\definecolor{palegreen3}{rgb}{0.484,0.801,0.484}
\definecolor{palegreen4}{rgb}{0.328,0.543,0.328}
\definecolor{paleturquoise}{rgb}{0.684,0.93,0.93}
\definecolor{paleturquoise1}{rgb}{0.73,0.996,0.996}
\definecolor{paleturquoise2}{rgb}{0.68,0.93,0.93}
\definecolor{paleturquoise3}{rgb}{0.586,0.801,0.801}
\definecolor{paleturquoise4}{rgb}{0.398,0.543,0.543}
\definecolor{palevioletred}{rgb}{0.855,0.438,0.574}
\definecolor{palevioletred1}{rgb}{0.996,0.508,0.668}
\definecolor{palevioletred2}{rgb}{0.93,0.473,0.621}
\definecolor{palevioletred3}{rgb}{0.801,0.406,0.535}
\definecolor{palevioletred4}{rgb}{0.543,0.277,0.363}
\definecolor{papayawhip}{rgb}{0.996,0.934,0.832}
\definecolor{peachpuff}{rgb}{0.996,0.852,0.723}
\definecolor{peachpuff1}{rgb}{0.996,0.852,0.723}
\definecolor{peachpuff2}{rgb}{0.93,0.793,0.676}
\definecolor{peachpuff3}{rgb}{0.801,0.684,0.582}
\definecolor{peachpuff4}{rgb}{0.543,0.465,0.395}
\definecolor{peru}{rgb}{0.801,0.52,0.246}
\definecolor{pink}{rgb}{0.996,0.75,0.793}
\definecolor{pink1}{rgb}{0.996,0.707,0.77}
\definecolor{pink2}{rgb}{0.93,0.66,0.719}
\definecolor{pink3}{rgb}{0.801,0.566,0.617}
\definecolor{pink4}{rgb}{0.543,0.387,0.422}
\definecolor{plum}{rgb}{0.863,0.625,0.863}
\definecolor{plum1}{rgb}{0.996,0.73,0.996}
\definecolor{plum2}{rgb}{0.93,0.68,0.93}
\definecolor{plum3}{rgb}{0.801,0.586,0.801}
\definecolor{plum4}{rgb}{0.543,0.398,0.543}
\definecolor{powderblue}{rgb}{0.688,0.875,0.898}
\definecolor{purple}{rgb}{0.625,0.125,0.938}
\definecolor{purple1}{rgb}{0.605,0.188,0.996}
\definecolor{purple2}{rgb}{0.566,0.172,0.93}
\definecolor{purple3}{rgb}{0.488,0.148,0.801}
\definecolor{purple4}{rgb}{0.332,0.102,0.543}
\definecolor{red}{rgb}{0.996,0,0}
\definecolor{red1}{rgb}{0.996,0,0}
\definecolor{red2}{rgb}{0.93,0,0}
\definecolor{red3}{rgb}{0.801,0,0}
\definecolor{red4}{rgb}{0.543,0,0}
\definecolor{rosybrown}{rgb}{0.734,0.559,0.559}
\definecolor{rosybrown1}{rgb}{0.996,0.754,0.754}
\definecolor{rosybrown2}{rgb}{0.93,0.703,0.703}
\definecolor{rosybrown3}{rgb}{0.801,0.605,0.605}
\definecolor{rosybrown4}{rgb}{0.543,0.41,0.41}
\definecolor{royalblue}{rgb}{0.254,0.41,0.879}
\definecolor{royalblue1}{rgb}{0.281,0.461,0.996}
\definecolor{royalblue2}{rgb}{0.262,0.43,0.93}
\definecolor{royalblue3}{rgb}{0.227,0.371,0.801}
\definecolor{royalblue4}{rgb}{0.152,0.25,0.543}
\definecolor{saddlebrown}{rgb}{0.543,0.27,0.0742}
\definecolor{salmon}{rgb}{0.977,0.5,0.445}
\definecolor{salmon1}{rgb}{0.996,0.547,0.41}
\definecolor{salmon2}{rgb}{0.93,0.508,0.383}
\definecolor{salmon3}{rgb}{0.801,0.438,0.328}
\definecolor{salmon4}{rgb}{0.543,0.297,0.223}
\definecolor{sandybrown}{rgb}{0.953,0.641,0.375}
\definecolor{seagreen}{rgb}{0.18,0.543,0.34}
\definecolor{seagreen1}{rgb}{0.328,0.996,0.621}
\definecolor{seagreen2}{rgb}{0.305,0.93,0.578}
\definecolor{seagreen3}{rgb}{0.262,0.801,0.5}
\definecolor{seagreen4}{rgb}{0.18,0.543,0.34}
\definecolor{seashell}{rgb}{0.996,0.957,0.93}
\definecolor{seashell1}{rgb}{0.996,0.957,0.93}
\definecolor{seashell2}{rgb}{0.93,0.895,0.867}
\definecolor{seashell3}{rgb}{0.801,0.77,0.746}
\definecolor{seashell4}{rgb}{0.543,0.523,0.508}
\definecolor{sienna}{rgb}{0.625,0.32,0.176}
\definecolor{sienna1}{rgb}{0.996,0.508,0.277}
\definecolor{sienna2}{rgb}{0.93,0.473,0.258}
\definecolor{sienna3}{rgb}{0.801,0.406,0.223}
\definecolor{sienna4}{rgb}{0.543,0.277,0.148}
\definecolor{skyblue}{rgb}{0.527,0.805,0.918}
\definecolor{skyblue1}{rgb}{0.527,0.805,0.996}
\definecolor{skyblue2}{rgb}{0.492,0.75,0.93}
\definecolor{skyblue3}{rgb}{0.422,0.648,0.801}
\definecolor{skyblue4}{rgb}{0.289,0.438,0.543}
\definecolor{slateblue}{rgb}{0.414,0.352,0.801}
\definecolor{slateblue1}{rgb}{0.512,0.434,0.996}
\definecolor{slateblue2}{rgb}{0.477,0.402,0.93}
\definecolor{slateblue3}{rgb}{0.41,0.348,0.801}
\definecolor{slateblue4}{rgb}{0.277,0.234,0.543}
\definecolor{slategray}{rgb}{0.438,0.5,0.562}
\definecolor{slategray1}{rgb}{0.773,0.883,0.996}
\definecolor{slategray2}{rgb}{0.723,0.824,0.93}
\definecolor{slategray3}{rgb}{0.621,0.711,0.801}
\definecolor{slategray4}{rgb}{0.422,0.48,0.543}
\definecolor{slategrey}{rgb}{0.438,0.5,0.562}
\definecolor{snow}{rgb}{0.996,0.977,0.977}
\definecolor{snow1}{rgb}{0.996,0.977,0.977}
\definecolor{snow2}{rgb}{0.93,0.91,0.91}
\definecolor{snow3}{rgb}{0.801,0.785,0.785}
\definecolor{snow4}{rgb}{0.543,0.535,0.535}
\definecolor{springgreen}{rgb}{0,0.996,0.496}
\definecolor{springgreen1}{rgb}{0,0.996,0.496}
\definecolor{springgreen2}{rgb}{0,0.93,0.461}
\definecolor{springgreen3}{rgb}{0,0.801,0.398}
\definecolor{springgreen4}{rgb}{0,0.543,0.27}
\definecolor{steelblue}{rgb}{0.273,0.508,0.703}
\definecolor{steelblue1}{rgb}{0.387,0.719,0.996}
\definecolor{steelblue2}{rgb}{0.359,0.672,0.93}
\definecolor{steelblue3}{rgb}{0.309,0.578,0.801}
\definecolor{steelblue4}{rgb}{0.211,0.391,0.543}
\definecolor{tan}{rgb}{0.82,0.703,0.547}
\definecolor{tan1}{rgb}{0.996,0.645,0.309}
\definecolor{tan2}{rgb}{0.93,0.602,0.285}
\definecolor{tan3}{rgb}{0.801,0.52,0.246}
\definecolor{tan4}{rgb}{0.543,0.352,0.168}
\definecolor{thistle}{rgb}{0.844,0.746,0.844}
\definecolor{thistle1}{rgb}{0.996,0.879,0.996}
\definecolor{thistle2}{rgb}{0.93,0.82,0.93}
\definecolor{thistle3}{rgb}{0.801,0.707,0.801}
\definecolor{thistle4}{rgb}{0.543,0.48,0.543}
\definecolor{tomato}{rgb}{0.996,0.387,0.277}
\definecolor{tomato1}{rgb}{0.996,0.387,0.277}
\definecolor{tomato2}{rgb}{0.93,0.359,0.258}
\definecolor{tomato3}{rgb}{0.801,0.309,0.223}
\definecolor{tomato4}{rgb}{0.543,0.211,0.148}
\definecolor{turquoise}{rgb}{0.25,0.875,0.812}
\definecolor{turquoise1}{rgb}{0,0.957,0.996}
\definecolor{turquoise2}{rgb}{0,0.895,0.93}
\definecolor{turquoise3}{rgb}{0,0.77,0.801}
\definecolor{turquoise4}{rgb}{0,0.523,0.543}
\definecolor{violet}{rgb}{0.93,0.508,0.93}
\definecolor{violetred}{rgb}{0.812,0.125,0.562}
\definecolor{violetred1}{rgb}{0.996,0.242,0.586}
\definecolor{violetred2}{rgb}{0.93,0.227,0.547}
\definecolor{violetred3}{rgb}{0.801,0.195,0.469}
\definecolor{violetred4}{rgb}{0.543,0.133,0.32}
\definecolor{wheat}{rgb}{0.957,0.867,0.699}
\definecolor{wheat1}{rgb}{0.996,0.902,0.727}
\definecolor{wheat2}{rgb}{0.93,0.844,0.68}
\definecolor{wheat3}{rgb}{0.801,0.727,0.586}
\definecolor{wheat4}{rgb}{0.543,0.492,0.398}
\definecolor{whitesmoke}{rgb}{0.957,0.957,0.957}
\definecolor{yellow}{rgb}{0.996,0.996,0}
\definecolor{yellow1}{rgb}{0.996,0.996,0}
\definecolor{yellow2}{rgb}{0.93,0.93,0}
\definecolor{yellow3}{rgb}{0.801,0.801,0}
\definecolor{yellow4}{rgb}{0.543,0.543,0}
\definecolor{yellowgreen}{rgb}{0.602,0.801,0.195}

\usepackage[numbers]{natbib}

\begin{document}

\title{Modeling and emulation of nonstationary Gaussian fields.}
\author{
Douglas Nychka\footnote{
Corresponding author:
Computational and Information Systems Laboratory, 
National Center for Atmospheric Research, PO Box 3000, Boulder, Colorado, USA, 80307-3000
}, \ 
 Dorit Hammerling\footnote{National Center for Atmospheric Research},
  Mitchell Krock\footnote{University of Colorado - Boulder}, and  Ashton Wiens\footnote{University of Colorado - Boulder} }

\maketitle

\begin{abstract}
Geophysical and other natural processes often exhibit non-stationary covariances and this feature is important to take into account for statistical models that attempt to emulate the physical process. A convolution-based  model is used to represent  non-stationary Gaussian processes that allows for variation in the correlation range and variance of the process across space. 
Application of this model has two steps: windowed estimates of the covariance function under the assumption of local stationary and encoding the local estimates into a single spatial process model that allows for efficient simulation. Specifically we give evidence to show that non-stationary covariance functions based on the Mat\`{e}rn family can be reproduced by the LatticeKrig model, a flexible, multi-resolution representation of Gaussian processes. We propose to fit locally stationary models based on the Mat\`{e}rn covariance and then
assemble these estimates into a single, global LatticeKrig model. One advantage of the LatticeKrig model is that it is efficient for simulating non-stationary fields even at  $10^5$ locations. This work is motivated by the interest in emulating spatial fields derived from numerical model simulations such as Earth system models. We successfully apply these ideas to emulate fields that describe the uncertainty in the pattern scaling of mean summer (JJA) surface temperature from a series of climate model experiments. This example is significant because it emulates tens of thousands of locations, typical in geophysical model fields, and leverages embarrassing parallel computation to speed up the local covariance fitting. 

\end{abstract}

{\ bf Keywords:}
Nonstationary Gaussian Process \  Markov Random Field \ Fixed Rank Kriging \  NCAR Large Ensemble Experiment


\section{Introduction}
In many areas of the geosciences it is natural to expect spatial
fields to be nonstationary. Not accounting for how the covariance
function may vary over space can result in misinterpreting the amount
of spatial correlations and also lead to unrealistic emulation of the
spatial fields. As spatial datasets grow in size and often have global
extent, it is more likely that one would expect nonstationary fields
simply because the spatial domain covers a heterogenous region. This
is often the case for surface climate fields where distinct land and
ocean regions might be expected to exhibit different spatial
structure.

Although large spatial data sets have the advantage of making it easier to
identify nonstationary covariances, they 
pose computational challenges when one attempts to apply standard
statistical models.  This feature is due to the well-known increase in
computational burden that grows as $O(n^3)$ where $n$ is the
number of spatial locations. Currently this feature effectively
prohibits fitting and simulating from Gaussian spatial process models
when the number of locations exceeds several thousand. Moreover, even for sample sizes where computation is still
feasible, interactive spatial data analysis will always benefits from
faster computation.

Given the spatial variation of a nonstationary covariance function it
is natural to focus on local modeling of the spatial field. Besides
reducing bias in the estimated covariance parameters, this strategy
also finesses some computational problems by converting a single large
problem into many smaller ones. A local approach does have the
disadvantage that it may not lead to a global model for the covariance
function or may imply a covariance model that is not readily
computed. This work combines efficient local covariance estimates with
a global model, LatticeKrig (LK, \cite{nychka2015multiresolution}), that can incorporate the local
information. The LK model is designed for statistical
computations for large data sets, and in particular it is possible to
simulate realizations from this model and make spatial predictions
with only modest computational resources. Although local covariance
estimates in aggregate require the same order of computation, the
memory demands are smaller and the computations can be easily done in
parallel. As a practical matter we exploit a large computing resource
(the NCAR supercomputer Cheyenne) for these computations and
find that the computation time scales almost linearly up 1000
processors (cores).

\begin{figure}[h]
\includegraphics[width=6in]{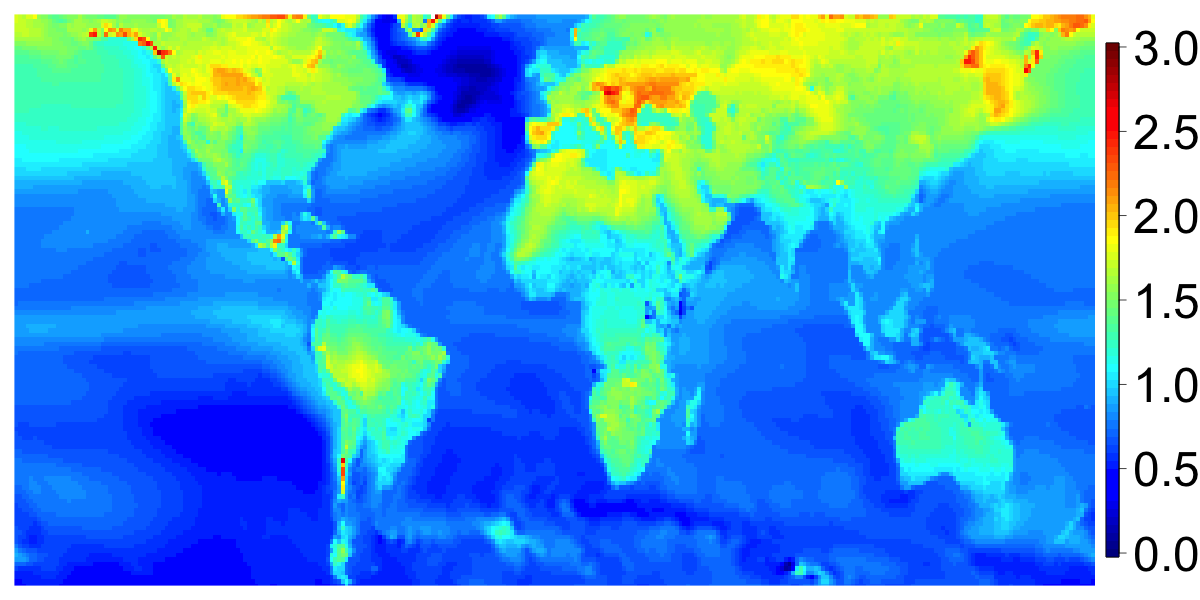}
\caption{Pattern scaling for mean surface temperature (Centigrade) for the 
months of June, July and August (JJA). This field is the sample mean 
across 30 ensemble members generated from the NCAR LEE during the 
simulation period 1920-2080 and uses the greenhouse gas scenario 
RCP8.5. The field is an estimate of the local response to global 
warming. For example a pixel value of 2.5 implies that that 1 degree
change in global average JJA temperature will result in a change of 2.5
at that location. }
\label{fig0}
\end{figure}

This work is motivated by a substantial example from impact assessment
modeling and Earth System science.  We have 30 derived fields from the
NCAR Large Ensemble Project (NCAR-LENS) \cite{kay2015community} that indicate 
the variation in local surface
temperature increase due to an increase in the global average.
The mean global pattern is illustrated in Figure \ref{fig0} and has the interpretation that a one degree change in global mean summer temperature will result in a change in local temperature according the values in this field. Such
fields form the basis of the pattern scaling technique in climate
science. One surprise from these different realizations in the NCAR-LENS is that there is significant variability about the mean scaling pattern in Figure \ref{fig0} (e.g. see bottom row Figure \ref{fig6A}). The data science goal then is to quantify this variability. 
This is a large spatial problem; the model grid is at
approximately one degree resolution and so there are more than 55,000
spatial locations ($288\times 192$ grid). Since these data cover the entire globe, even subdomains
 exhibit non-stationary behavior. Due the nature of
climate model ensemble simulations, one can assume that the 30 fields are independent
replicates from the same climate distribution. The goal is to model
these fields accurately and simulate additional realizations. A larger
set of realizations will be useful for quantifying the uncertainty of
impact assessment modeling of climate change.  Earth system models are large computer codes
 that can take months to run at dedicated supercomputing centers.  Strategies for extending the results using 
 fast statistical emulators is an important application to save additional computing resources. Moreover, detailed statistical models often reveal features of the simulations not obvious from basic data analysis. The specific spatial application in this
paper is part of a larger statistical emulation of surface temperature fields 
for extending 
model results to other conditions \cite{alexeeff2016emulating}. This application is
typical of climate model ensemble experiments and the availability of
replicated fields facilitates estimating non-stationary covariance
functions. 

The next section provides some background to this problem and presents
the convolution process model as a basis for considering non-stationary
covariances. Section 3 outlines the LK statistical model that is
useful for large spatial data sets and Section 4 gives evidence to
show that this model can approximate more standard covariance families
such as the Mat\'{e}rn. The application to the pattern scaling
ensemble is covered in Section 5 and we end with a discussion
section.

\section{Background}
We assume that the field of interest can be approximated as a Gaussian
spatial process, $y( \bbs )$, with $\bbs \in \cD \subset \Re^2 $ and
for convenience $\cD$ to be a rectangle. Furthermore, assume that this
field follows the additive model
\begin{equation} y(\bbs) = z(\bbs)^T\bbd +g(\bbs) + \epsilon(\bbs),
\label{obs.equ}
\end{equation}
where $z(\bbs)$ is a vector of covariates at each location, $\bbd$ a
vector of linear parameters, $g(\bbs)$ is a mean zero, smooth Gaussian
process, and $ \epsilon(\bbs)$ a Gaussian white noise process
independent of $g$.  The parameters $\bbd$ represent fixed effects in
this model while $g$ are $\epsilon$ are stochastic.

There are two features of the observational model which are
appropriate for our climate model application.  Let $\{ \bby^{m} \}$
index $M$ replicate fields that are independent realizations of the
additive model (\ref{obs.equ}). Given $N$ spatial locations $ \{
\bbs_1, \ldots, \bbs_N \} \subset \cD$, the observations are $Y_{i,m}
= y^{m}(\bbs_i) $, $M$ independent fields observed at $N$
locations. We also assume that the observations are complete -- every
replicate field is observed at all locations, which is typical for climate model output. 
Thus $\bbY$ can be
represented as an $N \times M$ matrix.  Besides assuming replicate
fields, we also assume that there is no measurement error in the
observations and the white noise process, $\epsilon$, is an
approximation to a fine scale process that is uncorrelated when
sampled on the scale of the observation locations.  In many
applications this is not the case and the underlying stochastic
process $g$ is the main component of interest. 

\subsection{Gaussian convolution process models}
Under the Gaussian process assumption, the distribution of $y$ is
determined by the covariance function for $g$ and the variance
function for $\epsilon$.  In particular we set
\[ E[ g(\bbs) g(\bbs^\prime) ] = k( \bbs, \bbs^\prime ) \mbox{ and }
E[ \epsilon(\bbs)^2 ]= \tau(\bbs). \]
Our main concern is to model $k$ without assuming stationary of the
process and to this end we use a convolution representation.

Let $ \psi$ be a continuous and square integrable function in $\Re^2$
and normalized so that
\[ \int_{\Re^2}\psi \left( || \bbu || \right) ^2 d \bbu = 1. \]
Define the spatially varying kernel function for {\it two} dimensions
as
\[ H(\bbs, \bbu) = \frac{1}{\theta(\bbs)} \psi \left( \frac{||\bbu -
\bbs||}{ \theta(\bbs)} \right),
  \]
and
\begin{equation} k( \bbs, \bbs^{\prime} ) = {\sigma(\bbs)}
{\sigma(\bbs^{\prime})} \int_{\Re^2} H( \bbs, \bbu) H( \bbs^\prime,
\bbu) d \bbu
\label{eq:convolution}
\end{equation}
where we assume that $\theta(\bbs)$ is at least  piecewise continuous 
and is interpreted as a range parameter varying over space.  Also
note that if $\sigma \equiv 1$ then $k$ is a correlation
function. Based on this form we see that $k$ will always be a valid
covariance function as it can be formally derived from the process
\[ g(\bbs) = \int_{\Re^2} H( \bbs, \bbu) dW(\bbu)  \]
with $dW(\bbu)$ a two dimensional standard, white noise process. 

The Mat\'{e}rn family is a popular choice for representing a covariance function and can
also be interpreted with respect to process convolution. Let 
\[ \psi(d) =  C(\nu) d^\nu K_\nu (d) \]
with $\nu > 0 $ a parameter controlling the smoothness of the process,
$K_\nu$ a modified Bessel function of the second kind, and $C$ a
constant depending on $\nu$. Assume that $\theta(\bbs) \equiv \theta$ and 
let  $H( \bbs, \bbu) =\psi(||\bbs-\bbu||/\theta)$.
Using the spectral representation of the Mat\'{e}rn it has been shown
\cite{zhu2010estimation} that $k( \bbs, \bbs^\prime)$ will also be a member of
the Mat\'{e}rn family with scale parameter still $\theta$. 
If $\nu$ is the smoothness for $g$, then $H$ must have  have smoothness
$\nu/2 - d/4$ .
For example, when $d=2$ and $\nu=2$, $g$ is obtained by
convolution using the exponential covariance ($\nu= 1/2$).  When the
scale parameter is not constant, however, the derived covariance is
not strictly Mat\'{e}rn and will not have a form that is readily
computed.

A convolution model to represent nonstationarity of a Gaussian process
has been addressed by many authors. In particular we highlight
early work in this area as applied to ocean temperature
data \cite{higdon1998process}, \cite{Higdon2002space} and subsequent
development \cite{fuentes_spectral}.  Although not
explicit, the more recent models  
 based on stochastic
partial differential equations  can also be tied to this kind of form
\cite{lindgren_2007},\cite{lindgren2011explicit}. If $H$ is the Green's function for a partial differential operator, $\cL$, then $g$ can also be identified with the solution: $ \cL g(\bbs) = dW(\bbs) $. An alternative to the convolution model is an explicit nonstationary covariance first proposed by Paciorek \cite{paciorek2006spatial} and extended 
to include smoothness parameters \cite{stein2005nonstationary}. Our understanding is that this model is derived as a scale mixture of Gaussian covariance convolutions
and so will not be the same as the direct convolution model sketched above. 

Some more recent work has addressed the computation for large
datasets \cite{zhu2010estimation} and using a low dimensional
function for the covariance parameters \cite{fuglstad_anisotropy}.
Recent work by \cite{fouedjio2016generalized} also is amenable to large data sets but focuses on the Paciorek form of covariance. A common thread in this past work is an emphasis on spatial prediction and not simulation of the unconditional, nonstationary process. Thus much of this work is not directly transferable to our application of statistical emulation.

To explain  the  algorithms for large data sets we review the relevant statistical computations
associated with Gaussian process inference. Although this work
considers maximum likelihood for inference, we note that the extension
to approximate Bayesian inference may also benefit from the
computational shortcuts that we highlight.  Let 
\[ K_{i,j}= \mathrm{Cov}(g(\bbs_i) , g(\bbs_j)),\]
and $R$ a diagonal matrix with elements
$R_{i,i} = \tau(\bbs_i)^2 $. This gives the covariance matrix for $\bby^{m}$ as $K +
R$. Also let $Z$ be a matrix with $N$ rows and each row being the
covariate vector $z(\bbs_i)$.  In vector/matrix form the likelihood
for the complete data set, $\bbY = [ \bby^{1}, \ldots,\bby^{M} $, is given by
\begin{equation}
\frac{N}{2} log(2\pi) - \frac{M}{2} log | K + R | -
\sum_{m=1}^M\frac{1}{2}(\bby^{m} -Z \bbd )^{T} ( K + R)^{-1} (\bby^{m}
- Z\bbd)
\label{eq:likelihood}
\end{equation}
with the covariance matrices implicitly depending on the
functions $\sigma$, $\tau$ and $\theta$.

To simulate a realization from this model one uses: $ \bby^* = A^T \bbe$
where $A^T A = ( K + R)$ and $\bbe$ a vector of i.i.d. $N(0,1)$
random variables. Typically one uses the Cholesky
factorization to obtain $A$, although we mention some benefits of using
the symmetric square root for an approximate simulation algorithm
below.

For large data sets evaluating the likelihood poses the well-known
computational hurdles of storing $K$, solving the linear system
associated with $( K + R)$, evaluating the determinant of $K +R$ and
also finding a square root for simulation.  In addition, for a
non-stationary model, evaluating the covariance as a convolution may
also involve significant computation if the integral does not have a
closed form.

These features make it difficult to estimate non-stationary
models. Here we take a local approach by assuming that the covariance
function is approximately stationary in a small spatial neighborhood
and we take the stationary parameter estimates for $\tau$, $\sigma$
and $\theta$ as representing the values of these parameter surfaces in
the center of the neighborhood.  This is not a new idea and has roots 
dating back to the early work on moving window Kriging 
\cite{haas1990_acid}, \cite{haas1990kriging}, \cite{VerHoef2004} and is
also similar to local likelihood ideas \cite{stein2004approximating}
\cite{stein2011}.
\subsection{Local unconditional simulation}
This local strategy can also be extended to a simple algorithm for
simulating a non-stationary process and  outline this algorithm for
comparison with a global approach given later.  Suppose
one seeks to simulate a realization of the process for a set of
locations. Generate i.i.d. $N(0,1)$ random variables for these locations
and combine them in a vector $\bbv$.  Now for a grid point, $\bbs^*$
evaluate the stationary covariance matrix using $\sigma( \bbs^*)$,
$\tau(\bbs^*)$, and $\theta(\bbs^*)$ for all locations in a
neighborhood of $\bbs^*$.  Denote this covariance matrix as $\Sigma^*$ and let
$\Sigma^{* 1/2}$ be its symmetric square root. Let $\bbv^*$ be the
subset of $\bbv$ that corresponds to the neighborhood of $\bbs^*$ and
let $\bbw^*$ be the row of $ \Sigma^{* 1/2}$ that corresponds to
$\bbv^*$ (or the neighborhood of $\bbs^*$).  The simulated value for the field at $\bbs^*$ 
is $(\bbw^*)^T\bbv^*$. Repeat these steps for all other locations using
the {\it same} realization of the white noise vector.  Keeping the
white noise vector the same creates the spatial dependence in the
field. Moreover, if the points are on a fine grid one can make a
heuristic argument that this is a local and discrete approximation to
the convolution process. We base this connection on the idea that the  
 symmetric square root of
a covariance matrix will be an approximation to the kernel $H$.  Finally we
note that the use of a symmetric square root seems appropriate over using an 
upper triangular decomposition. A moving window based on an lower triangular weighting of neighboring 
locations will depend on the ordering of locations and may produce artifacts.
For example, if the center grid point happens to be the first row in this matrix it will depend on just a single
component of $\bbv$.

In contrast to local simulation described above, what is new in this
current work is a global covariance model that assembles the local
estimates into a single process description. It is possible to
simulate from this model exactly for a large number of spatial locations and 
without making local restrictions.  Also new
is the use of a highly parallel computing strategy to find the local
maximum likelihood estimates of covariance parameters. 

\section{LatticeKrig model}

The process convolution model (\ref{eq:convolution}) is a useful
nonstationary model but difficult to implement for large spatial
data. Here we present an alternative,  the LK
model  that is a good approximation to
standard covariance families but is much more amenable to fast
computation.  

The LK model is one of several recent approaches 
to handle large spatial data in a consistent global way. The recent review 
\cite{heaton2017methods} compares many of these methods 
with an emphasis on spatial prediction for a data set of $10^5$ locations.   An important consideration is that fields such as the climate model example
can have small nugget variances ($\tau^2$) and so the representation needs to have adequate
degrees of freedom to represent the process at fine resolution. Some methods based on low-rank basis functions
 may not have this feature. Another consideration is that the method
 must admit a global process representation and be able to simulate a Gaussian process efficiently.  
 The multi-resolution approximation \cite{katzfuss2017multi}, hierarchical nearest neighbor methods 
 \cite{datta2016hierarchical} and stochastic partial differential equation models \cite{lindgren2011explicit}
 might all be alternatives to using LK for the global simulation. 
 
The basic idea  of LK is to adopt fixed rank Kriging  (FRK, see
\cite{cressie2008fixed}, \cite{katzfuss2011}) but model the {\it
precision} matrix of the coefficients as a sparse matrix. Ironically FRK was first 
developed as a low dimensional approach to large data sets. The LK model, on the other hand by exploiting 
sparse matrix methods, can handle a large number of basis functions 
 comparable to the number of spatial
locations. 
This model draws on the work of FRK and stochastic PDEs
 but also adds a multi-resolution elaboration that greatly improves
its flexibility. Moreover, the non-stationary LK model can be
interpreted as a superposition of convolution-type processes at
different spatial scales. 

We assume that the process, $g(\bbs)$, is the sum of $L$ independent latent
processes
\begin{equation}
\label{equ:multires}
g(\bbs) =   \sum_{l=1}^{L}   g_l( \bbs).
 \end{equation} 
 Here $g_l(\bbs)$ has mean zero and marginal variance $\sigma_l
(\bbs)^2 $. Thus, the marginal variance for $g$ is $\sum_{l=1}^L
\sigma_l (\bbs)^2.$
 
\subsection{Multi-resolution basis}

Each component, $g_l$ is defined through a basis function expansion
as
\begin{equation} g_l( \bbs) = \sum_{k=1}^{m(l)} c_k^l
\varphi_{k,l}(\bbs),
\label{process.equ}
 \end{equation}
 where $\varphi_{k,l}$, $1\le k \le m(l)$, is a sequence of fixed
basis functions and $\bbc^l$ is a vector of coefficients distributed
multivariate normal with mean zero and covariance matrix,
$\bbQ^{-1}_l$. Coefficients are assumed to be independent between the
different levels.

We now rewrite the observational additive model under the LK process.
Stack the coefficient vectors for the different levels, $\bbc = \{ \bbc_1, \ldots,
\bbc_L \}$.  Let $\Phi^{l}_{i,j} =
\varphi_{j,l}(\bbs_i)$, be a matrix where rows index the observation
locations, columns index the $m(l)$ basis functions at the $l^{th}$
level and combine these matrices into a single matrix  $\Phi = [ \Phi^1, \ldots, \Phi^L] $. With these aggregations, the additive model for an
observed field is
 \begin{equation} \bby^{m} = Z\bbd + \Phi \bbc^{m} + \epsilon^m ,
\label{eq:LKobs}
\end{equation}
where $m$ denotes the $m^{th}$ replicate. 

 To achieve a multi-resolution, the basis functions are formed from
translations and scalings of a single radial function.  Let $\phi$ be
a unimodal, symmetric function in one dimension, and for this work we
assume that it is compactly supported on the interval $[-1,1]$.  The
basis functions depend on a sequence of nested rectangular grids, $\{
\bbu_{j,l} \}$, $ 1\le j \le m(l)$ and where $1 \le l \le L$.  The
grid spacing is kept at the same distance in both dimensions and
decreases by a factor of 2 from $l$ to $l+1$.  Thus, in two dimensions
$m(l)$ increases approximately as the exponential function $4^{l}$.

Consistent with radial basis function terminology, we will refer to
the grid points as {\it node points}.  We adopt a scale parameter
$\delta$ to set the overlap of the basis functions and the basis
functions are then defined as
\begin{equation}
 \varphi_j^* = \phi( 2^l ||\bbs- \bbu_j||/ \delta).
\label{RBF.equ}
\end{equation}
Note that the power of 2 scaling means the basis functions at
each level have the same overlap.  Indeed, except for scaling and
translation they all have the same shape.  Here the $^*$ indicates
that these are not exactly the final versions of the basis functions
but will be normalized as described in the Appendix. Although an
important detail for implementation, the normalization is not crucial
for understanding the main features of this model.

\subsection*{Spatial Autoregression}
In this application the spatial covariance for $\bbc^l$ is modeled as
a non-stationary Markov random field. The coefficient vector $\bbc_l$
at a single level follows a spatial autoregression (SAR) and is
organized by the node points.  That is, each coefficient, $c_{k,l}$,
is associated with a node point $u_{k,l}$ and  
and will have up to four nearest neighbors. Denote
this set $\cN_k$. We assume that for a sequence of parameters $a_{k}$
and $v_{k,l}$ i.i.d N(0,1) random variables
\begin{equation}
 a_{k} c_{k,l} - \sum_{k^* \in \cN_k} c_{k^*,l}  = v_{k,l},  
\label{SAR}
\end{equation}
where $a_{k} > 4$ for the process to be stable.  Let $B_l$ be the SAR
matrix that is square with the same dimension as $ \bbc_l $. The
diagonal elements of $B_l$ are $a_{k}$. In the isotropic case, the off
diagonal elements are $-1$ at the positions of the nearest neighbors
and the remaining entries are zero.  With this construction $B_l\bbc_l
= \bbv_l $, and simple linearity implies that the precision matrix for
$\bbc_l$ is $Q_l= (B_l^T B_l)$. $Q_l$ are also the weights one
associates with $\bbc_l$ as a Gaussian Markov random field (GMRF). If
the GMRF is specified directly then $Q_l$ must be constrained to be
positive definite. By constructing $Q$ via the SAR, however, one is
guaranteed this condition will hold. For constant $a$ parameter this
GMRF has been studied in \cite{lindgren2011explicit} and approximates a Mat\'{e}rn
covariance with smoothness $\nu=1.0$ and range parameter given by
$\kappa = 1/\sqrt{a-4} $. 
\subsection{An approximate convolution process}
We can also conjecture how this model behaves as a 
discretized convolution process.  A realization of $g_l$ at the observations has the representation
\begin{equation}
\bbg_l  = \Phi_l  B^{-1} \bbv_l
\end{equation}
where the matrix multiplications in this expression are  sums over the lattice points. Given that the lattice
is equally spaced and the support of the basis functions is calibrated to overlap several lattice points, this 
expression may approximate integrals over the spatial domain. From the discussion in 
\cite{zhu2010estimation}  (see Table 1) $ B^{-1}$ can be associated with a  Mat\'{e}rn kernel with smoothness $0$ and is denoted    as $K_0$ in \cite{zhu2010estimation}.  
We conjecture that that a limiting argument should associate $\Phi_l  B^{-1}$ as proportional to 
\[ 
\int_{\Re^2} \phi( 2^l||\bbx -\bbw||/ \delta ) K_0(||\bbw- \bbu ||)  d\bbw.
 \]
Although $K_0$ has a singularity at $0$, convolution with the Wendland basis functions used in this work are smooth at zero and  will result in a bounded kernel $H$.

\subsection*{Computational efficiency}

 To simulate from the LK model it is
enough to simulate a realization of the coefficients since the basis is fixed.  We use the
assumption that levels are assumed to be independent and let
$\bbc^*_l$ be a realization of the coefficients at the $l^{th}$ level.
\begin{equation} Q_l^{-1} \bbc^*_l = \bbv
\label{eq:simc}
\end{equation}
where $\bbv$ are iid N(0,1) random variables. $g_l$ is now evaluated
using \ref{process.equ} and the components are added.  $B_l$ is a sparse matrix with at most 5 nonzero entries
per row. Thus $Q_l$ will also be sparse with at most  13 nonzero
entries.  Evaluating $Q_l^{-1} \bbv$ is efficient because one can
solve the linear system in (\ref{eq:simc}) using a sparse Cholesky
decomposition of $Q_l$. Moreover, $\Phi_l$
will also be a sparse matrix  due to the compact support of the basis functions and so evaluating $g_l$ is also
efficient.

The log likelihood for the LK model is also based on (\ref{eq:likelihood}) where
$K= \Phi Q^{-1} \Phi^T$. Using the Sherman-Morrison-Woodbury formula
the quadratic form in the likelihood can be computed efficiently based
on the sparsity of the matrices $R$, $\Phi$ and $Q$. Exploiting
Sylvester's identity, the determinant can be evaluated efficiently as
well. Finally, we note the concentration of the likelihood by
substituting in the estimates for the $\bbd$, which is equivalent to finding
a generalized least squares (GLS) estimate with covariance $(K + R)$. This GLS
estimate can also be found easily using the same techniques for
evaluating the  quadratic form in the likelihood.

\subsection*{Properites of the LK covariance}
\begin{figure}[h]
\includegraphics[width=3in]{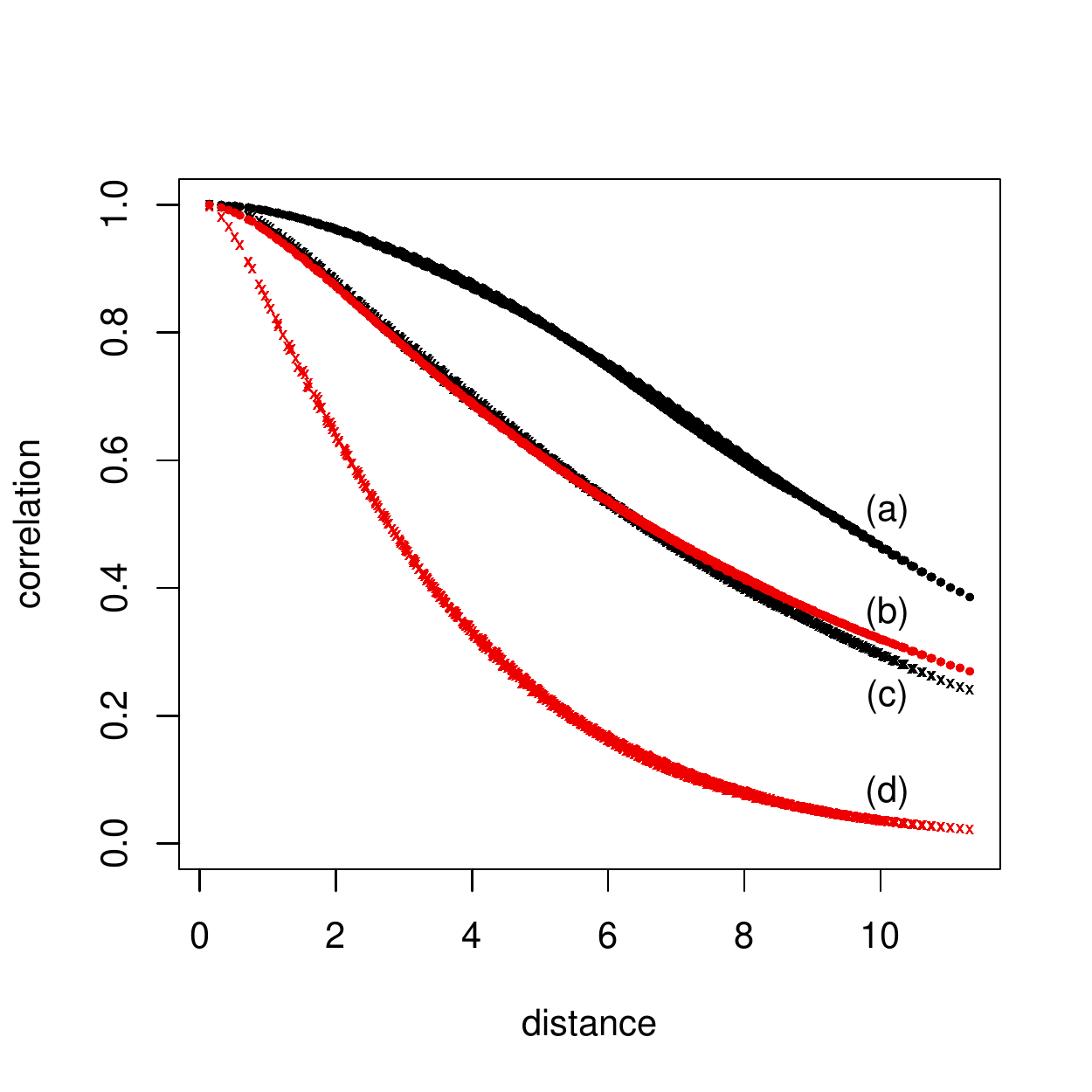}
\
\includegraphics[width=3in]{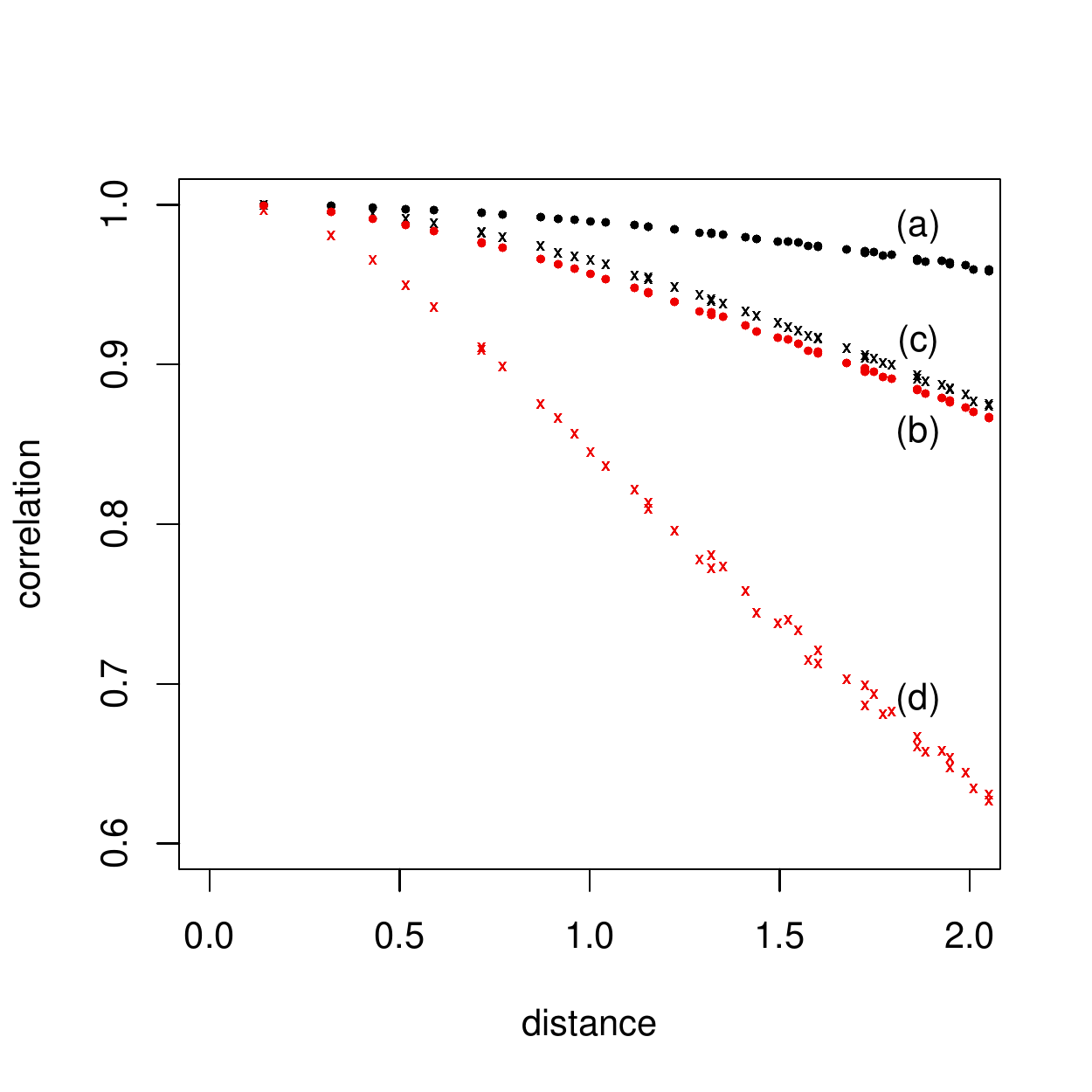}
\caption{Examples of varying covariance parameters in the LatticeKrig
model. Plotted are  correlation functions between a grid of points
in the square domain $[-8,8]\times[-8,8]$ and the location $(0,0)$.
The left hand plot is an enlargement of the right side one to show the
similarity of the (b) and (c) covariance models.}
\label{fig1}
\end{figure}

The convolution covariance can be understood as a locally stationary
process and so it is also useful to interpret the LK model from a
stationary perspective. The stationary version just omits the $k$
subscripts and dependence on $\bbs$. That is $a_{k,l} \equiv a_l$ and
$\sigma_l(\bbs) \equiv \sigma_l $.  Figure \ref{fig1} is an example of
correlation functions for a four level LK model with a Wendland
radial basis function (see Appendix) and an overlap of 2.5 in grid spacing units.
The plot illustrates the effect of the $\sigma_l$ and $a$ parameters.
The spatial domain is taken to be $[-8,8] \times [-8,8]$ and the
initial grid spacing is $4$.  The covariance is nearly stationary and
so the correlation function is plotted as a function of the distance
between locations in the domain and the center, $(0,0)$.  The slight spread in the
points from a fixed curve is the result of the slight departures of
the LK model from being exactly isotropic. The black points (a) represent a covariance with
$a$ = 5 and only weights at the coarsest level, $(1,0,0,0)$. The  black ``x" points (c) indicate a covariance that also
has $a$ = 5 but sets $\sigma_1^2, \sigma_2^2, \sigma_3^2, \sigma_4^2$
proportional to the weights $(1, 0.5, 0.25, 0)$. The red covariance (d)
also uses $a$ = 5  and the weight sequence $(0, 1, 0.5, 0.25)$, eliminating a
contribution from the coarsest level. Finally the covariance (b) uses
this sequence but sets $a$ = 4.1 to give a much longer correlation
range.
The curves (b) and (c) are derived from different levels of resolution but still have very similar behavior. This indicates one pitfall of the LK model in that the parameters across multiple levels may not be well identified for a given covariance function. 

 The theory in \cite{LatticeKrig} proves an asymptotic result that indicates
the LK model can approximate the smoothness of members of the
Mat\'{e}rn family as the number of levels becomes infinite.  The main
finding is that $\sigma_l$ should be chosen to decay as $2^{-l\nu}$ to
approximate a Mat\'{e}rn process with smoothness $\nu$.  To build the
best approximation over a limited number of resolution levels,
however, it is more accurate to optimize the LK parameters
numerically. In this work we use three levels and target the climate model
application. Following the local simulation of the field described in
Section 2 we think it is appropriate to optimize the mean squared
error for the weights ($\bbw^*$) used to simulate the central field
value.   We find that for $\theta$ in the interval $[1,12]$ and for a smoothness of $\nu=1.0$ 
the LK model can approximate the
Mat\'{e}rn to within a few percent of relative root mean squared
error. We obtain approximations with less than 6 percent relative root mean squared error for the case $\nu=2.0$ over the interval $[1,8]$. We expect the approximation to improve if additional levels are added. 
\subsection{Non-stationary version of the LK model}
The LK model can be made nonstationary in two ways:  spatial variation in  the
weighting across scales, $\{ \sigma_l \}$ and also spatial variation in the parameter $a$. The overall
correlation range is controlled by $\sigma_l(\bbx)$ and can select
among the scales of the basis functions.  The parameters $a_k$ control
the dependence of the GMRF and adjust the correlation range of $g_l$
within a level. It is possible to fit the LK model directly, however,
 fully unconstrained parameters may not be identifiable. 
This property is apparent  in Figure \ref{fig1} where covariance functions using two distinct levels (b and c) 
are quite similar. 
 In
this work we take a parsimonious approach of estimating a local
Mat\'{e}rn model and then encoding the LK parameters to approximate
that model. This strategy has a secondary benefit that a standard
covariance model is faster to estimate locally for small numbers of
spatial locations. 
We use two informative test cases to explore the
properties of the LK approximation. The spatial domain is taken to be
$[-24,24] \times [-24, 24]$. For the first case and we divide the region vertically into two
different correlation ranges:
\[ 
\theta(\bbs) =  \left\{
\begin{array}{cc}
                                5 & \mbox{ for }   \bbs_1  \le 0 \\
                                1.9 &  \mbox{ for }  \bbs_1 \ge 0
\end{array}
\right.
\]
 while fixing $\sigma(\bbx) \equiv 1 $ and $\tau(\bbx) \equiv 0 $.

\begin{figure}
\includegraphics[width=6in]{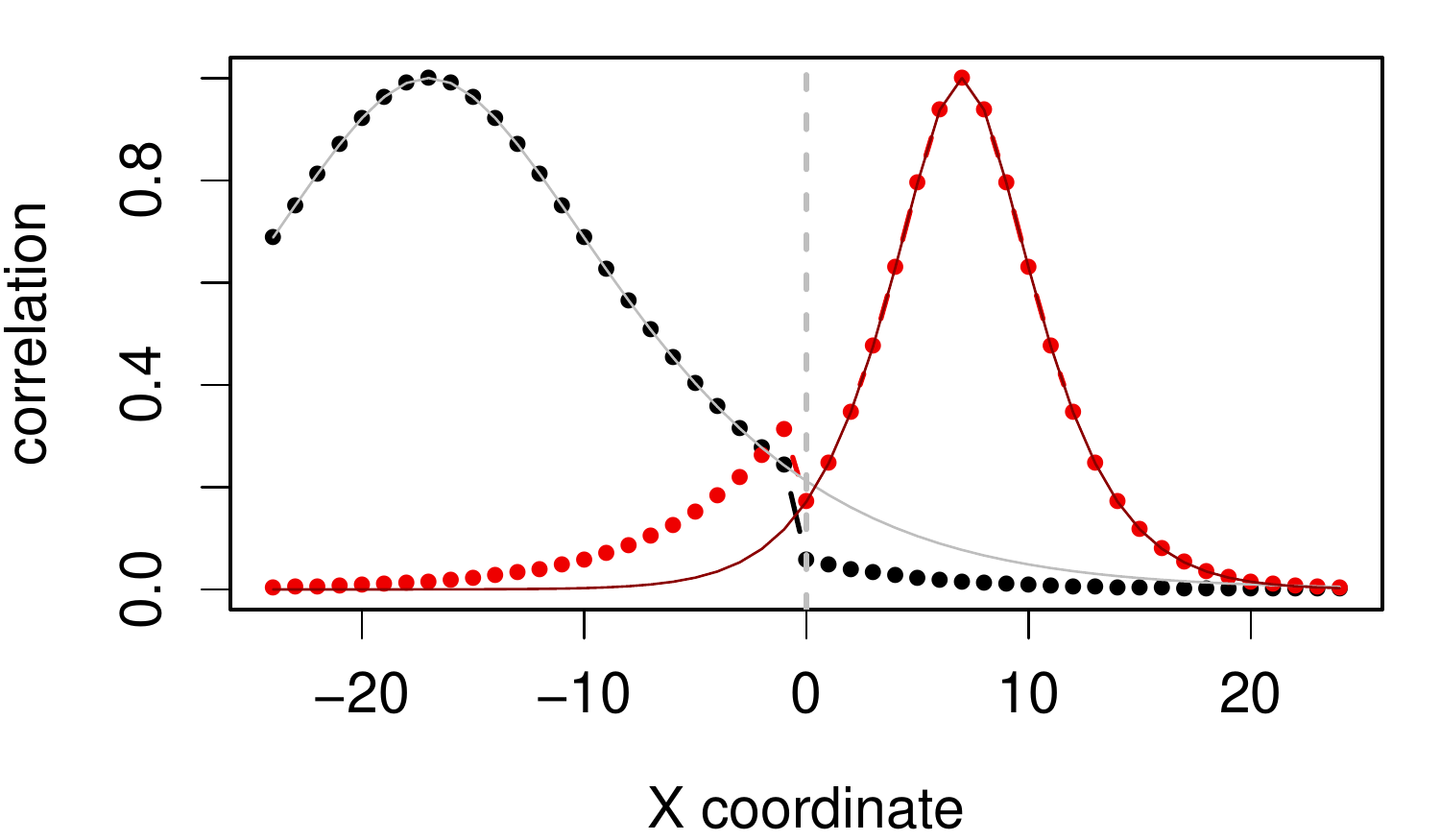}
\caption{ Correlation curves illustrating the  non-stationary first test case. Superimposed are the local stationary correlation functions. The spatial domain for this example is the square $[-24, 24]\times[-24, 24]$ but the correlation function is evaluated is evaluated along the transect with the Y-coordinate equal to zero. Plotted are the correlation functions for  the location $(-17,0)$ in black and $(7,0)$ in red with points. The grey lines are the stationary correlation functions using the range parameter at these locations.  
 }
\label{fig2}
\end{figure}

\begin{figure}
\includegraphics[width=6in]{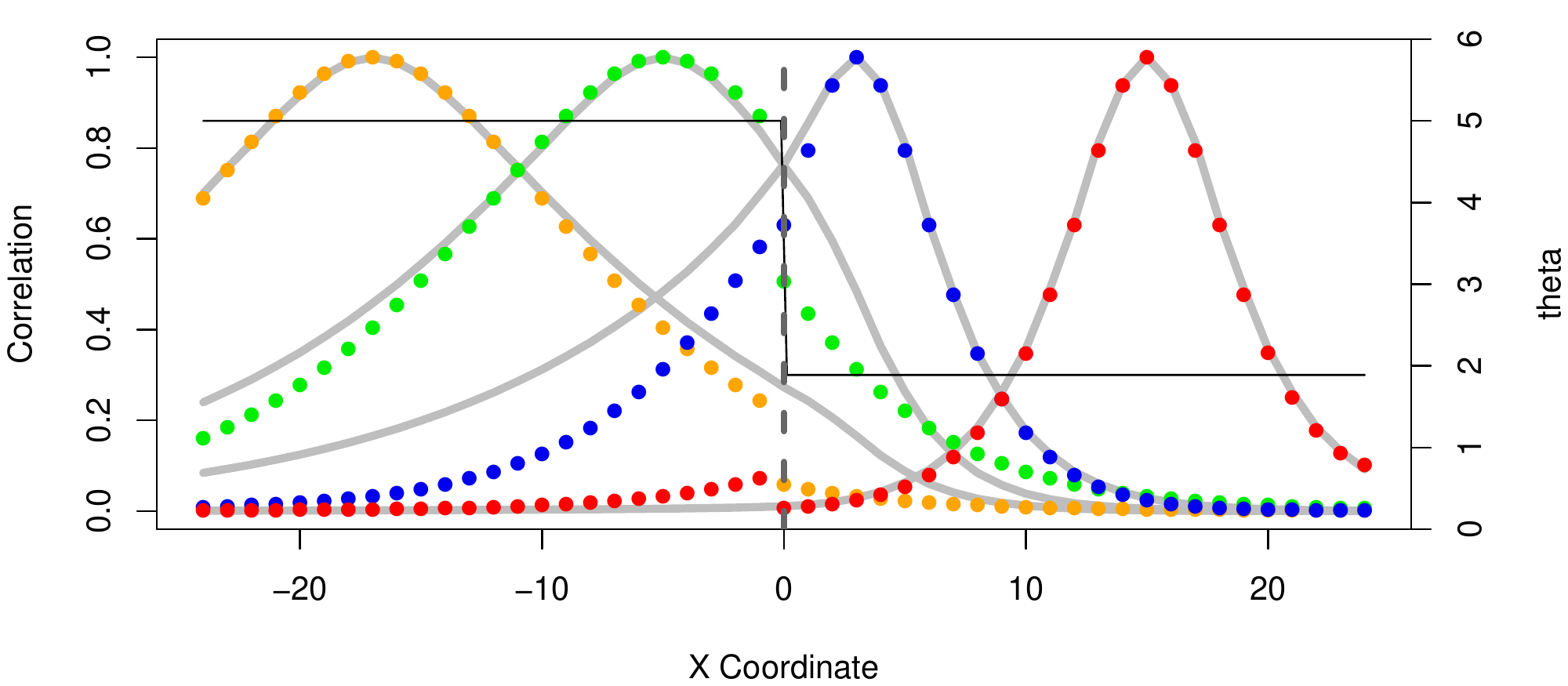}
\caption{Comparison of LatticeKrig approximation (lines) and true
non-stationary correlations (points) for the first non-stationary test case, a discontinuous range 
parameter. The superimposed black line gives the values for 
$\theta(\bbs)$ as a function of the X-coordinate and corresponds to 
the axis 
on the right hand side of the plot. The correlation functions are with
respect to the locations  $(-17,0)$,  $(-5,0)$, 
$(3,0)$, and  $(15,0)$.
 }
\label{fig3}
\end{figure}

 Figures \ref{fig2}  and \ref{fig3} summarize the success of this kind of encoding.
The non-stationary field was defined by this range parameter and
convolving exponential kernels according to (\ref{eq:convolution}). Based on the
properties of the Mat\'{e}rn we expect a stationary Mat\'{e}rn
covariance function with smoothness $2$ when $\theta$ is
constant. (The discrete set of points used for plotting was to limit
the size of the computation.)  Figure \ref{fig2}  illustrates the correlation
function of the field at two locations along the the Y-coordinate at 0.
Restricting to a horizontal transect was done to limit 
the amount of computation. For reference, superimposed are the
Mat\'{e}rn covariance functions assuming local stationarity. Note that
these tend to track the non-stationary curves except at the boundary
where $\theta$ changes. Also the note surprising lack of montonicity
in the correlation function for $(7,0)$ .  Figure \ref{fig3} reproduces these true 
non-stationary correlation functions and superimposes the correlation
functions from the LK approximation. Here the LK model is encoded to
be a locally stationary Mat\'{e}rn approximation with a spatially
varying $a$ parameter. The precise value of $a$ is found by interpolating $\theta(\bbs)$ to the node points and then converting $\theta$ to $a$ using the stationary approxmation.  Overall the LK model
appears to capture the general features of the non-stationarity and
the transition from $\theta = 5$ to $1.9$ at $\bbx_1 = 0$. The LK model makes a smoother transition, however, across this boundary tending to over estimate
correlations for locations closer to the discontinuity. The LK non-stationary model also misses the departure from monotonicity in the
correlation function. The second non-stationary test case is setup similar to the first except that $\theta(\bbs)$ is taken to vary as a linear function in $\bbs_1$ decreasing from $6$ at the left boundary to $1$ at the right. Figure \ref{fig3} compares the true correlation functions to those approximated by the LK model. In this case the agreement is good and we attribute this to the smoothly varying choice for the $\theta(\bbs)$ field. 

The final figure is a realization of the LK
approximation for the first test case and gives a qualitative impression of the variation in
the correlation scale across the discontinuity in $\theta(\bbx)$ (grey vertical line). The two previous plots only depict the correlation along the
$\bbs_2 =0 $ transect and this is indicated by the black line in this
figure.  To simulate the true field in this first case would not be
difficult because $\theta(\bbs)$ is piecewise constant. In general the
simulation would be computationally intensive, requiring a separate
convolution kernel computation for each location in the field. Even if the non-stationary matrix could be  assembled there is still the standard challenge to find the Cholesky factor  for a large  and dense covariance matrix.  In contrast, the LK
realization is on a $129\times129$ grid and took under 20 seconds to compute on a MacBook Air laptop (Intel Core i7, 2.2 GHz, 8 GB memory) using serial code in R.  
  
 \section{Simulating variation in pattern scaling fields} 
 
 As outlined in the introduction our application is to model the
spatial variation among the patterns derived from the NCAR-LENS project.  We
will focus on a North and South America sub-domain to streamline
presentation comprising  $13056 = 102 \times 128$ grid boxes.  We found
that a Mat\'{e}rn with $\nu = 1.0$ was a reasonable choice for
smoothness across the domain and an isotropic Mat\'{e}rn covariance was fit locally
using several sizes of moving square windows.  Here we report
estimates based on $11 \times 11$ pixel windows with the maximum
likelihood estimates registered to the center grid box. 

\subsection{Local covariance estimates}
Even with 30 replicate fields, estimating the $\theta$ and $\sigma$
parameters was not robust and we often obtained very large values over
the ocean.  This sensitivity is expected from large correlation
ranges but we note that $\tau$ is still adequately estimated and is
small over the ocean reflecting a smooth spatial field. Let $
\sigma_{obs}$ be the sample standard deviation for the replicates and
for each grid box.  A simple adjustment to $\sigma$ for these cases is
for a threshold of $\hat{\tau} < .003 $ and $ \hat{\sigma} >
\sigma_{obs} $ we take take $ \hat{\sigma} \equiv \sigma_{obs} $. For
$\hat{\theta} > 15 $ we set $\theta \equiv 15$.  Admittedly these are
crude adjustments but they respect the basic assumptions of more
spatial coherence over the ocean and also the fact that correlation
ranges beyond $15$ degrees ($> 1600$ km at the equator) are not likely
or will not influence the simulation of the fields. Previous work (e.g. \cite{zhu2010estimation},
\cite{fouedjio2016generalized}) has considered the local covariance estimates 
as under-smoothed and applied a second smoothing step to the estimated parameter. 
We applied
an approximate thin plate spline (the function {\tt fastTps} from the fields package \cite{fields})
with the smoothing parameter found by maximum likelihood
to the log of the estimated $\theta$ field. Here the tapering radius  was set to 10 degrees so a moderate number of locations were included in the smoothing. 
 The smoothing parameter indicated little additional 
 smoothing. Given 13056 observations the effective degrees of freedom for the spline was over 3500 and a surface plot confirms this impression. We also fit a thin plate spline model with the land/ocean mask added as a linear covariate and this did not change the results. Given this data analysis we concluded that there was little benefit in adding a second modeling step in representing the range parameter field. 
 
\subsection{Simulation of the pattern scaling uncertainty}
Figure  \ref{fig5} reports the Mat\'{e}rn estimates based on the above
discussion. Perhaps the most important aspect of these data is the
striking non-stationarity in all three parameter fields and the clear
land/ocean demarcations along much of the coast line. We believe that this clear 
signal on land/ocean in the parameters suggests that our choice of window size is 
appropriate and overall the parameters are being estimated in a robust way. 
The higher
variability ($\sigma$) in the spatial process ($g$) in the center of
North America and over the land area near Argentina is reasonable, along with a larger white noise component ($\tau$) over land. Although
not shown, the ratio of white noise to smooth process variance 
($\tau^2/\sigma^2$) tends to
be larger over land.

The implementation of the LK model is available in the R LatticeKrig package \cite{LatticeKrig}.
These parameters were encoded into a LK model with three levels of
resolution where the coarse grid spacing is 2.5 degrees. The fields were simulated on the 
grid of the model, roughly 13K locations, although the LK algorithm does not require locations on a grid and 
simulation took under 60 seconds on a MacBook Air system. Almost all of that time was in setting up the matrices $\Phi$ and the Cholesky decompositions of $\{ Q_l \}$ and there is little overhead for generating more than one realization.  The reader should 
reconcile this timing with the method of local simulation that requires an eigen decomposition at each grid point. In this case 
the simulation algorithm will take on the order of hundreds of minutes as a serial computation for a $9\times9$ window. Of course this can be parallelized in the same way as the local parameter estimation, however, the simulations will be local with the simulated process being independent when windows/weights do not overlap. 

Figure  \ref{fig6A} 
shows four realizations of the LK process on the top row, and for
reference the first four ensemble members from the spatial data set
are given on the bottom row. Qualitatively the spatial coherence and
variability matches between these simulated and true cases. We note
the emulation does have some modest deficiencies. For example, the anisotropy  over
the Equitorial Pacific is not well represented. In the model appears to be longer correlation scales in the East-West direction as compared to the North-South.  Of course, this is not a failing
of the LK approximation but rather the use of an isotropic covariance
function. As a contrast to the non-stationary model we also generated 
stationary realizations. The top row of Figure \ref{fig6B} gives four realizations of a stationary field 
  using the  median of the parameter estimates  over land. The bottom row is the same except the medians over the   ocean are used. The similarity between top and bottom plot is deliberate; to aid in this comparison we use the white noise vectors for generating the land and ocean field in each column. The differences between these two choices of stationary models are striking and it is clear that neither would provide an accurate emulation of the model output.
  
\subsection{Parallel implementation}
This example was computed using a parallel strategy and the R language \cite{team2000r}. Fitting the
spatial model for each window is an embarrassingly parallel operation
and moreover the ensemble data set fields are relatively  small (about
12Mb). We took the approach of using a supervisor R session and then
spawning many R worker sessions. The supervisor session assigns tasks  (i.e. specific local windows) to
each worker based on balancing the work load.  When a worker is done
with a specific task the information from the fitting is passed back the supervisor.  
The complete set of results are assembled as an output list in the supervisor session and in our
case this output list has as many components as grid boxes in the spatial domain
and each contains the results of the local fitting. Creating the
workers, broadcasting the spatial data set, and assigning the tasks is
all done in R through the {\tt Rmpi} package \cite{rmpi}. In using R we have leveaged the stable and rich set of spatial analysis tools that are available to the community. In particular the maximum likelihood estimates are found using the {\tt spatialProcess} function from the fields package and is called in exactly the same way as on a laptop.  We have used this
approach on the NCAR supercomputer {\it Cheyenne}  \cite{cheyenne} and found it exhibits excellent (strong) scaling. An example of timing is given below in Figure \ref{fig7}.
In this test case a one level LK model limited to a 1000 grid boxes is fit directly to the data rather than
the Mat\'{e}rn covariance. Here we
see linear scaling in the time with up to 1000 parallel R worker
sessions. As expected the time to spawn workers shows a linear increase
(orange points)
but is an order of magnitude smaller that the time spent in computation (blue points). Note that this
scaling has attractive practical implications. Using 1000 cores will
result in nearly a factor of 1000 speedup in the analysis and can potentially convert a lengthly batch analysis into one that is almost interactive. 
 
 \section{Discussion}
 Combining local covariance estimation with a global model provides a practical route for 
 modeling and simulating large spatial data sets. 
 We have shown that the LK model can reproduce abrupt non-stationarity
  in a process where the range parameter has a discontinuity and as expected does well when the range parameter varies smoothly across a spatial domain.  Moreover, in places where the process is locally stationary we see that there is close agreement between the Mat\`{e}rn correlation function and the approximate one from the LK 
  representation. The advantage of the LK representation is the ability to generate unconditional realization of the process at large number of locations. One can also use the LK model for spatial prediction and inference \cite{heaton2017methods} although that role is not needed for climate model emulation.
  
 Most data analysis represents a compromise between model complexity and realism and the need to estimate the model accurately from data. In this work we focused on data that has spatial replicates and this makes parameter estimation much more stable. In addition we do not believe this data set to be an isolated example as ensemble climate experiments are now the norm in climate science. The local covariance models could be improved by adding anisotropy and also covariates for the land/ocean regions. Because the LK likelihood can be evaluated for the complete data set, there is the opportunity to fit parameters that have a global extent, such as land/ocean effects, along with local covariance parameters. Some parameters such as anisotropy could also have larger spatial extent in order to provide stable estimate.  The estimation strategy in this case  would have the flavor of back-fitting in additive models where one would alternate between fitting different components of the model until convergence is obtained. 
 
 Local covariance fitting has the advantage that it leverages standard spatial tools and diagnostics. A more comprehensive approach would be to fit the fields of parameter maximizing the complete likelihood. To this end the local estimates could be used as the initial values in  a LK model and subsequent optimization can refine these parameters. Recent work in deep learning has benefited from the use of stochastic gradient descent for fitting many parameters in an artificial neural  network and this technique may also be useful for optimizing the parameters here. 
 
 From an analytical perspective it would be useful to determine the differences between the explicit non-stationary  models following  Paciorek's construction and those derived via process convolution.  Preliminary results, not reported here, suggest that these models are substantially different for the case of a discontinuity in the range parameter. In general we have found a process based description to be more fruitful for arriving at covariance models. For example, we build connection weights in a SAR model rather than the more general MRF framework. There is an additional benefit from process models that they can admit physical interpretation and be more  useful to domain scientists.  
 
 The use of embarrassingly parallel steps, such as local covariance fitting or local simulation, is a computational strategy that merits more attention. Here we have developed code mainly in R to manage this process and so this framework is accessible to any accomplished R user. Indeed, the framework we use on the supercomputing system is the same that we use on a laptop except for several lines of batch scripting and changing directory pathnames. We also believe that this style of computation may drive alternative models and algorithms as the number of processors/cores available for routine spatial data analysis grows. 
 
\section*{Acknowledgements}
This work was supported in part by the National Center for Atmospheric Research (NCAR) and also the National Science Foundation Award 1406536. NCAR is sponsored by the National Science Foundation and managed by the University Corporation for Atmospheric Research. We also acknowledge high-performance computing support from Cheyenne (doi:10.5065/D6RX99HX) provided by NCAR's Computational and Information Systems Laboratory, sponsored by the National Science Foundation.

\appendix

\section{Wendland radial basis kernel}
 The
Wendland functions are compacted supported on $[0,1]$ and   are also positive
definite. Below is the version of the Wendland  valid up to 3
dimensions and belonging to $C^4$:
\[
 \phi(d) = \left\{ \begin{array}{cl} 
 (1-d)^6 ( 35d^2 + 18d +3)/3 & \mbox{ for  }  0 \le d  \le 1 \\ 
                   0                          & \mbox{ otherwise}.
                  \end{array}
                \right.
\]
  
\section{Normalization to approximate stationarity}
Because of the discrete nature of the SAR the marginal variance of the
LatticeKrig process will not be constant in the spatial domain. This can cause artifacts in the estimated surface and compromised its ability to approximate stationary covariance functions. To adjust the marginal variances we compute the unnormalized variance and divide by this quantity to give a constant variance at any location. 

 Based on the model and notation from Section 3, let $\bbC= \bbQ_l^{-1} $ and at multi-resolution level $l$,  
\[
\mathrm{Var}( g_l(\bbs)) =  \sum_{j,k} \varphi_{j,l}(\bbs)\bbC_ {j,k} \varphi_{k,l}(\bbs)
\]
  Accordingly, let $\omega(\bbs) = \sqrt{\mathrm{Var}(g_l( \bbs))}$ 
  and normalize the basis functions as
   \[ \varphi_{j,l}(\bbs) =   \sigma_l(\bbs) \frac{\varphi_{j,l}^*(\bbs)}{ \omega(\bbs)}  \] 
   These are the actual basis functions used in the spatial analysis. 

\bibliography{BRACE}
\bibliographystyle{plain}

\section*{Figures}

\begin{figure}
\includegraphics[width=6in]{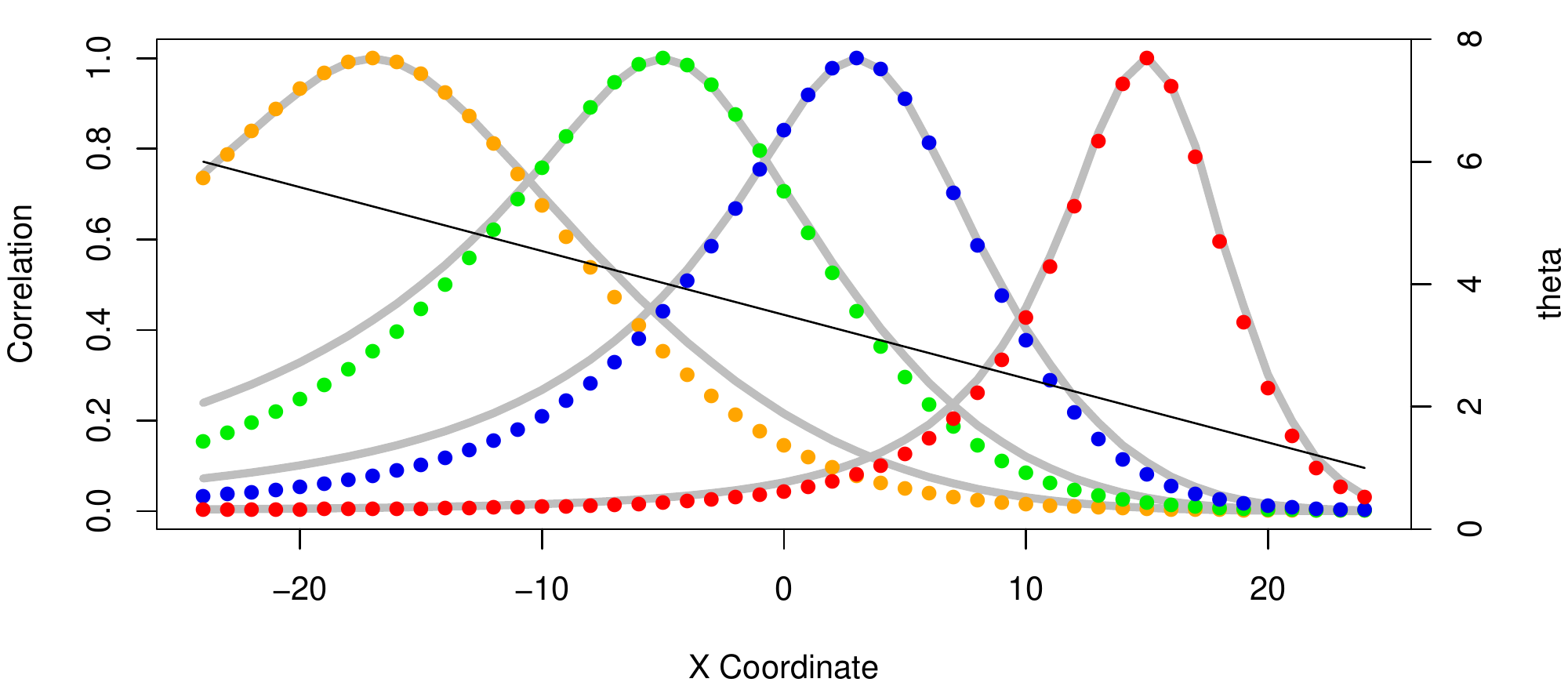}
\caption{Comparison of LatticeKrig approximation (lines)  and true  non-stationary correlations (points) for the second non-stationary case, a linearly varying range parameter. As in figure \ref{fig3} the black line indicates the value of the range parameter.  
 }
\label{fig3B}

\end{figure}

\begin{figure}
\includegraphics[width=6in]{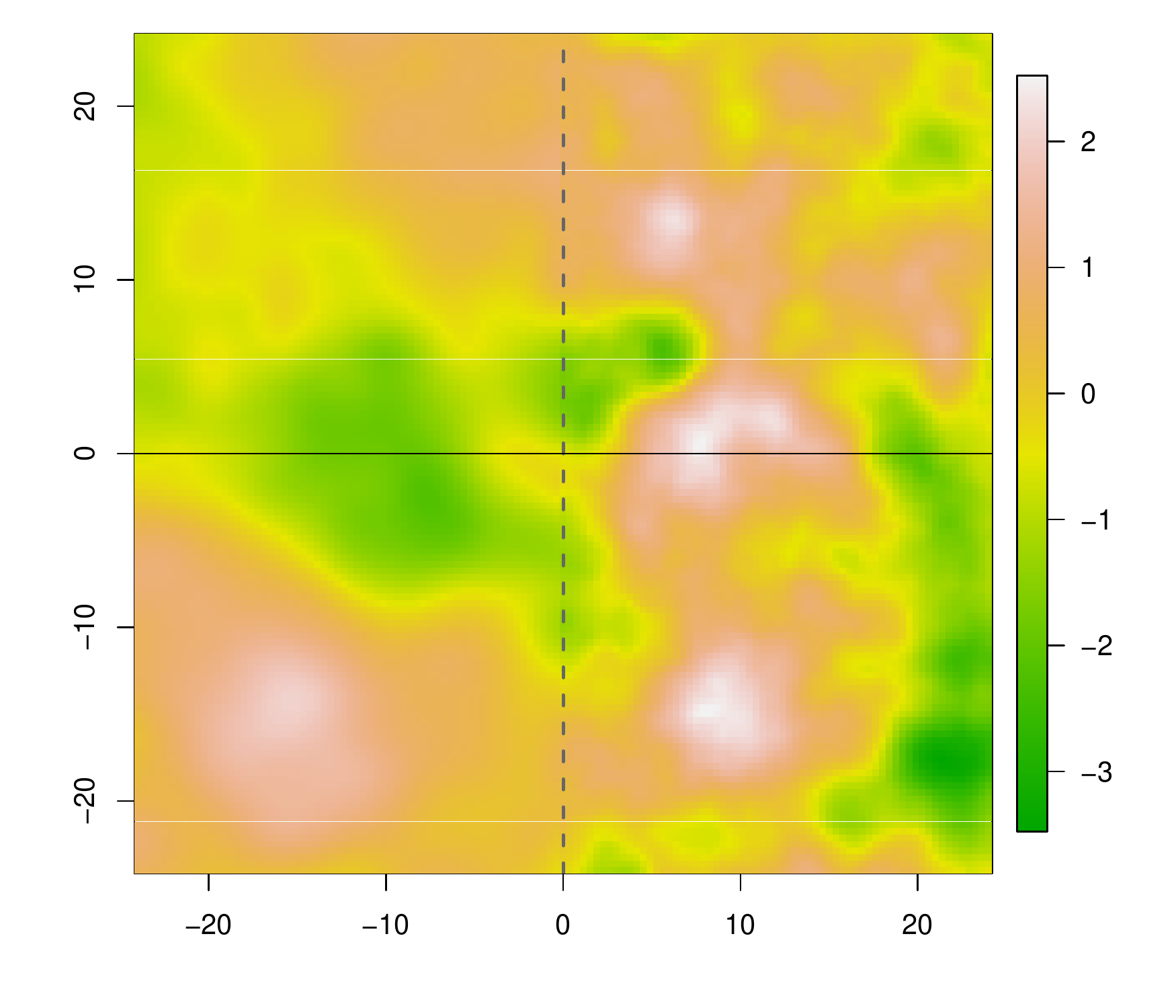}
\caption{Simulated field from the  LatticeKrig approximation from the first non-stationary case. The vertical line is where the range parameter changes from $5$ to $1.9$. The horizontal lines is the transect used to evaluate the correlation functions in the previous figures. 
 }
\label{fig4}
\end{figure}

\begin{figure}
\includegraphics[width=6in]{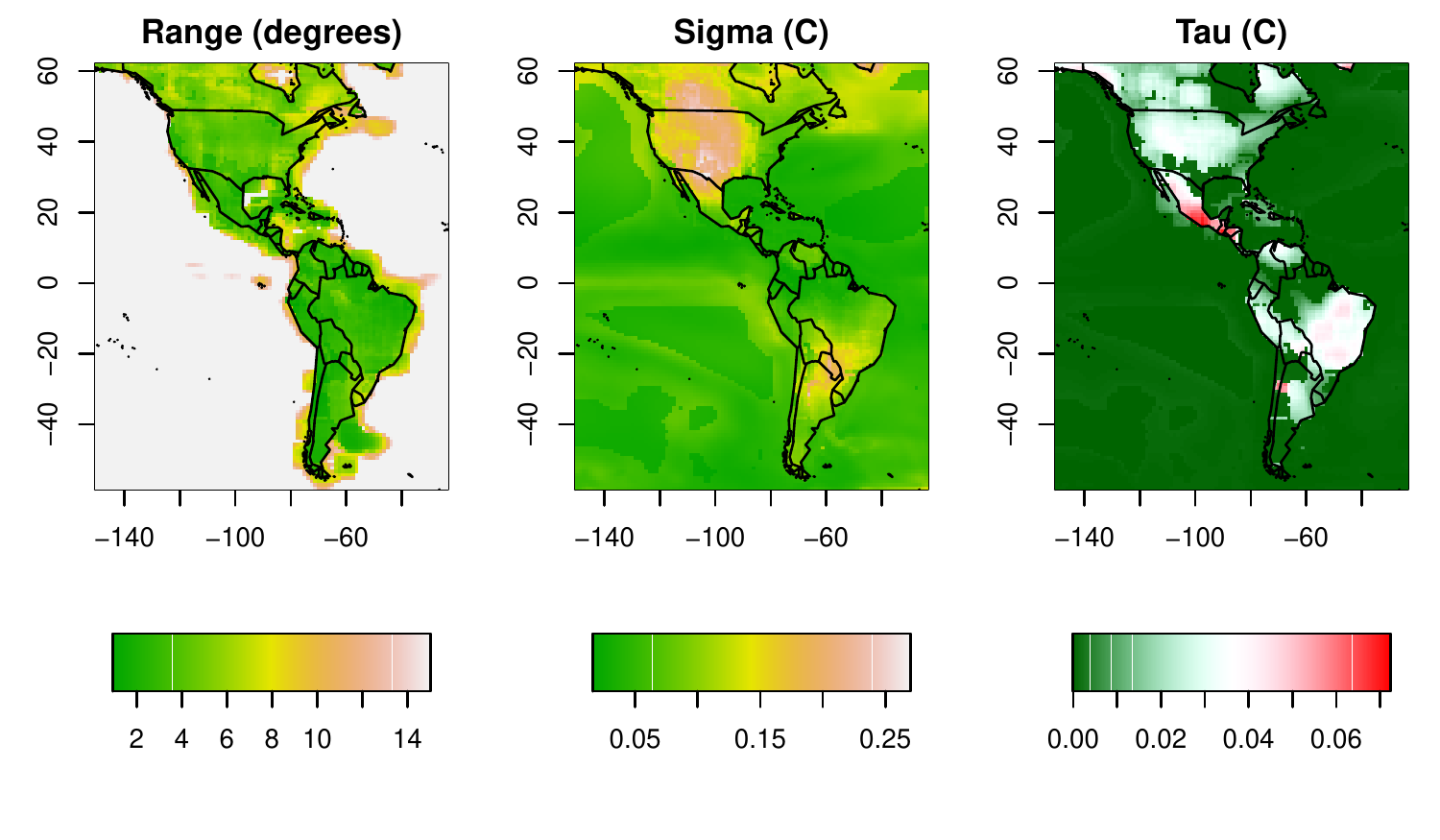}
\caption{Mat\'{e}rn parameter fields based on a $11\times11$ pixel moving window. At the equator this window width is  13.75 degrees or 1526 km. Parameters are found by maximum likelihood in these local windows but the $\sigma$ and $\theta$ fields have been truncated for large values over the ocean. 
 }
\label{fig5}
\end{figure}

\begin{figure}
\includegraphics[width=6in]{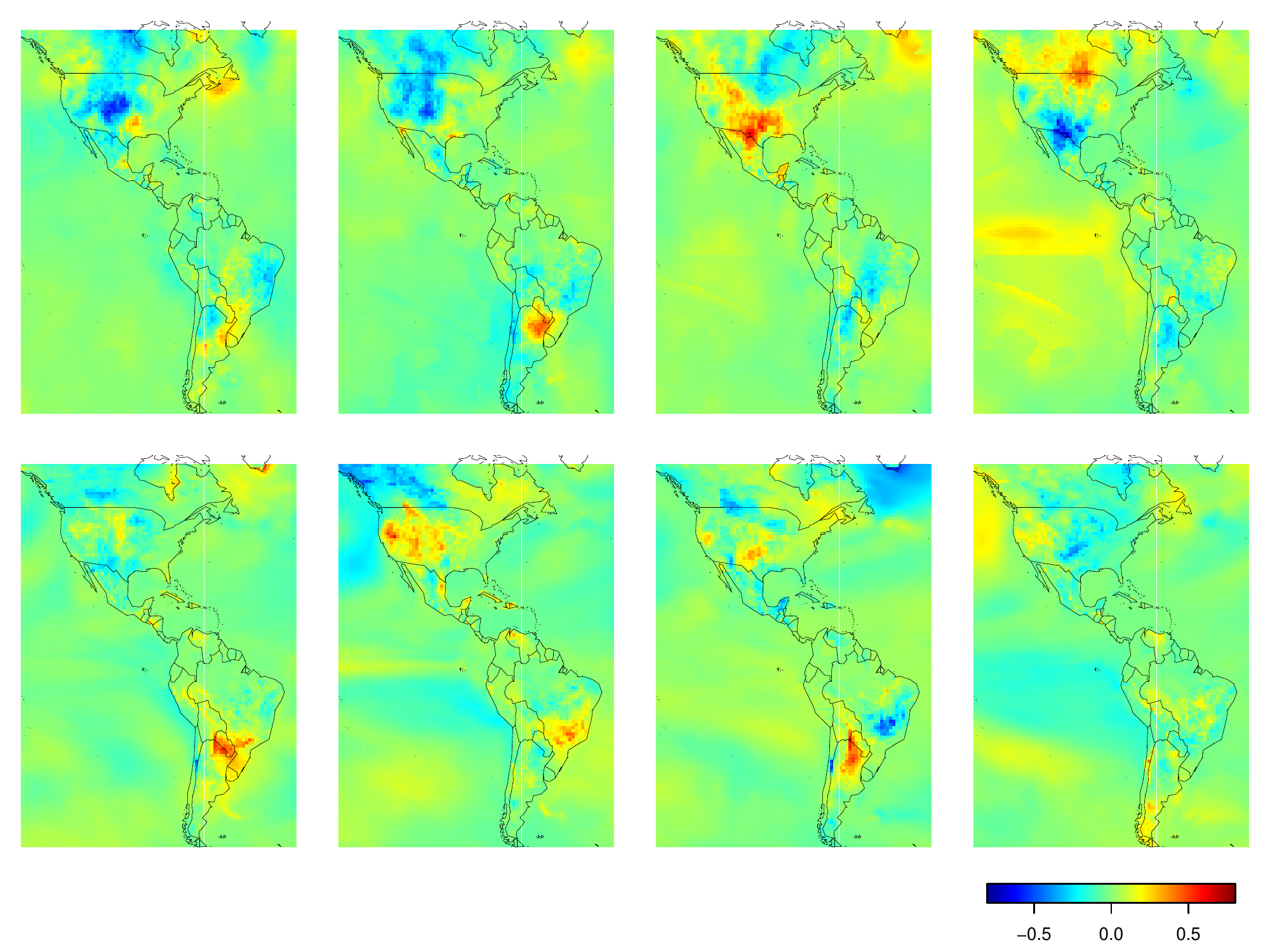}
\caption{ Simulated and true fields for the pattern scaling data set. Top row are four realizations from the LK Gaussian process model and the bottom plots are the 
first four data fields based on the climate model output.}
\label{fig6A}
\end{figure}
\begin{figure}
\includegraphics[width=6in]{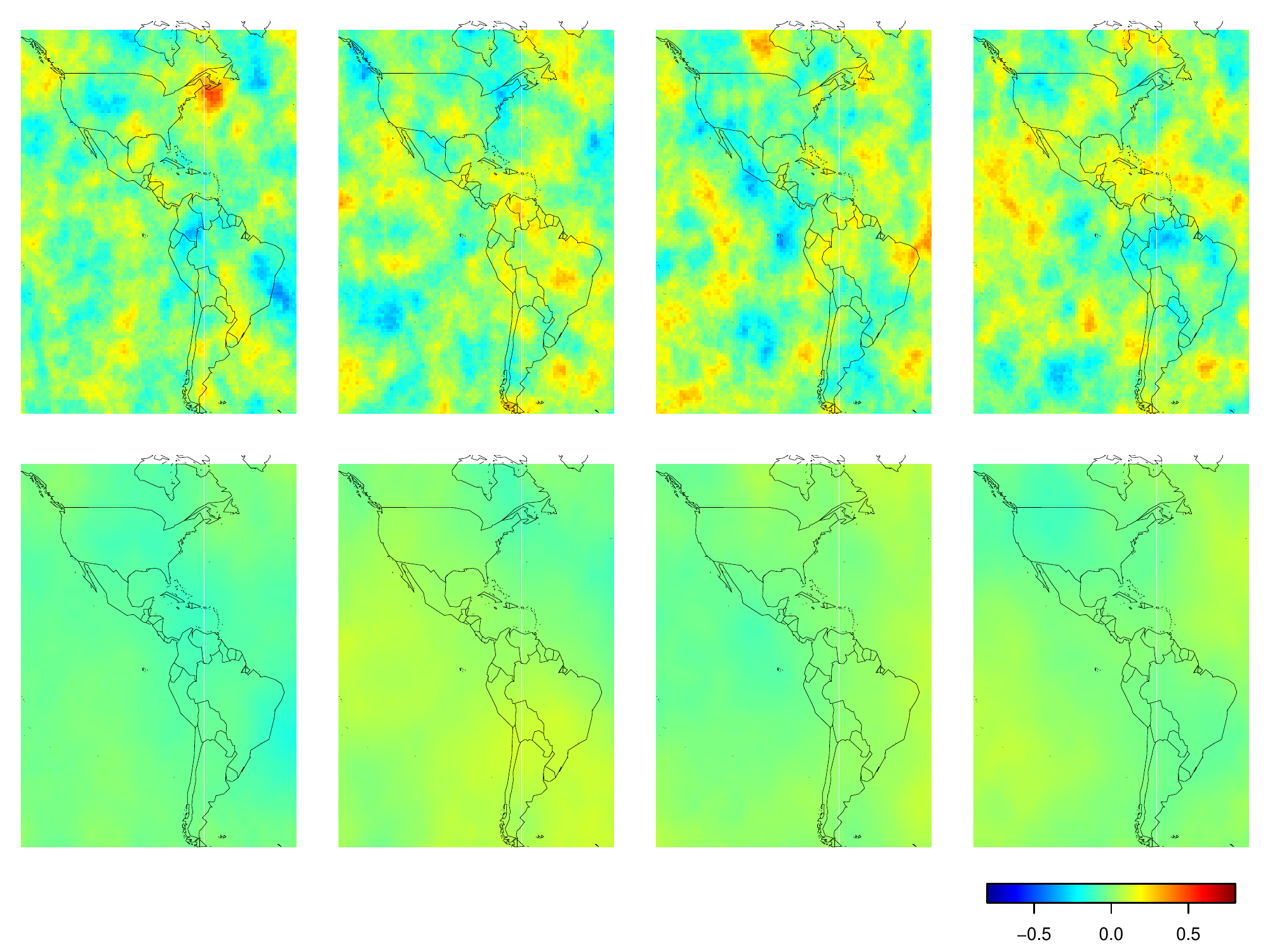}
\caption{ Simulated stationary fields following the pattern scaling data set. Top row are four realizations from the LK Gaussian process model using the median covariance parameters over land. The bottom row are  the corresponding realizations using median parameters from over ocean. }
\label{fig6B}
\end{figure}

\begin{figure}
\includegraphics[width=6in]{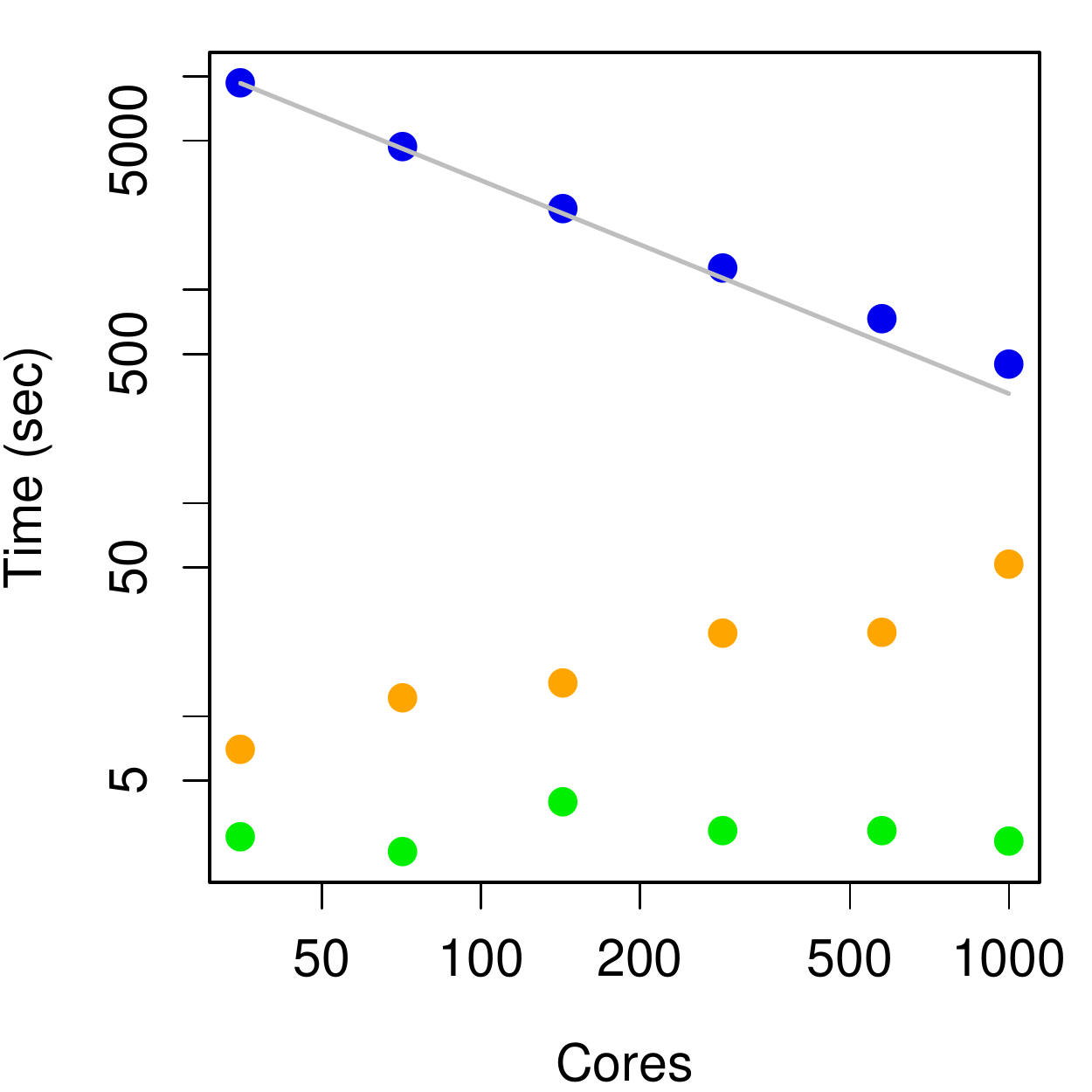}
\caption{Timing results for fitting local stationary covariances to 1000 grid boxes as a function of the cores. In this case the number of cores is equal to the number of worker R sessions. The parallel sessions were managed by the Rmpi package and done on the Cheyenne supercomputer managed by NCAR. 
 Blue- time spent fitting model, green - time  to broadcast data to workers and orange - time to spawn workers.}
\label{fig7}
\end{figure}

\end{document}